\documentclass[lettersize,journal]{IEEEtran}
\usepackage{amsmath,amsfonts}
\usepackage{xcolor}
\usepackage{algorithmic}
\usepackage{algorithm}
\usepackage{array}
\usepackage[caption=false,font=normalsize,labelfont=sf,textfont=sf]{subfig}
\usepackage{textcomp}
\usepackage{stfloats}
\usepackage{url}
\usepackage{hyperref}
\usepackage{verbatim}
\usepackage{graphicx}
\usepackage{cite}
\usepackage{adjustbox}
\usepackage{multirow}
\usepackage{tablefootnote}
\usepackage{threeparttable}
\usepackage{booktabs}
\usepackage{amssymb}
\usepackage{balance}
\definecolor{positive}{rgb}{0,0.5,0}
\definecolor{negative}{rgb}{0.5,0,0}
\hypersetup{
  colorlinks,
  linkcolor={red},
  citecolor={blue!50!black},
  urlcolor={blue!80!black},
}

\begin{document}

\title{Transmit What You Need: Task-Adaptive Semantic Communications for Visual Information}

\author{Jeonghun Park,~\IEEEmembership{Student Member,~IEEE,} and Sung Whan Yoon,~\IEEEmembership{Member,~IEEE,}
        % <-this % stops a space

\thanks{Manuscript received November 16, 2024; revised May 1, 2025. This work was supported by the Institute of Information \& communications Technology Planning \& Evaluation (IITP) grant funded by the Korea government (MSIT) (No. RS2020-II201336, Artificial Intelligence Graduate School Program (UNIST)), (No. RS-2025-25442824, AI Star Fellowship Program at
UNIST), and (No. IITP-2024-RS-2022-00156361, Innovative Human Resource Development for Local Intellectualization program), and by the National Research Foundation of Korea (NRF)
grant funded by the Korea government (MSIT) (RS-2024-00459023), and by a grant of the Korea Health Technology R\&D Project through the Korea Health Industry Development Institute (KHIDI), funded by the Ministry of Health \& Welfare, Republic of Korea (grant number: RS-2025-02223382).}

\thanks{Jeonghun Park and Sung Whan Yoon are with the Graduate School of Artificial Intelligence, Ulsan National Institute of Science and Technology (UNIST), Ulsan 44919, Republic of Korea (e-mail: {jhpark2024, shyoon8}@unist.ac.kr). Sung Whan Yoon is also with the Department of Electrical Engineering, UNIST. Sung Whan Yoon is the corresponding author.}

\thanks{© 2025 IEEE. This is the author’s version of the article accepted for publication in IEEE Journal on Selected Areas in Communications. The final version of record is available at: https://doi.org/10.1109/JSAC.2025.3623159
}
% <-this % stops a space
}

\markboth{Accepted to IEEE Journal on Selected Areas in Communications (Early Access). DOI:10.1109/JSAC.2025.3623159}%
{J. Park \MakeLowercase{\textit{et al.}}: Transmit What You Need: Task-Adaptive Semantic Communications for Visual Information}

%\IEEEpubidadjcol

\bstctlcite{BSTcontrol}

\maketitle

\begin{abstract}
Recently, semantic communications have drawn great attention as the groundbreaking concept surpasses the limited capacity of Shannon's theory. 
Specifically, semantic communications are likely to become crucial in realizing visual tasks that demand massive network traffic.
Although highly distinctive forms of visual semantics exist for computer vision tasks, a thorough investigation of what visual semantics can be transmitted in time and which one is required for completing different visual tasks has not yet been reported. To this end, we first scrutinize the achievable throughput in transmitting existing visual semantics through the limited wireless communication bandwidth. In addition, we further demonstrate the resulting performance of various visual tasks for each visual semantic. Based on the empirical testing, we suggest a task-adaptive selection of visual semantics is crucial for real-time semantic communications for visual tasks, where we transmit basic semantics (e.g., objects in the given image) for simple visual tasks, such as classification, and richer semantics (e.g., scene graphs) for complex tasks, such as image regeneration. 
To further improve transmission efficiency, we suggest a filtering method for scene graphs, which drops redundant information in the scene graph, thus allowing the sending of essential semantics for completing the given task. We confirm the efficacy of our task-adaptive semantic communication approach through extensive simulations in wireless channels, showing more than 45 times larger throughput over a naive transmission of original data.
Our work can be reproduced at the following source codes: \hyperlink{https://github.com/jhpark2024/jhpark.github.io}{https://github.com/jhpark2024/jhpark.github.io}
\end{abstract}

\section{Introduction}
\IEEEPARstart{A} recent surge in visual applications, which demands substantial traffic, such as virtual and augmented reality, has exacerbated the disparity between the increasing volume of transmitted data and the limited communication capacity \cite{ref1, ref2, ref3}. 
In this circumstance, the \textit{semantic communication} paradigm has attracted significant interest as a key enabler to break the limit of communication capacity \cite{ref4, ref5, ref6, ref7}.
In the mid-20$^{\text{th}}$ century, the concept of semantic communication was originated by extending Shannon's classical communication theory to the categorized levels of communications, which acknowledged the required transmission accuracy for the different levels of settings \cite{ref8}.
While Shannon's original theory primarily focuses on the so-called \textit{technical level}, which prioritizes the accurate transmission of raw bits over a channel, the higher level setting so-called \textit{semantic level} emphasizes not merely the transmission of bits but the delivery of the compressed semantics of the source data, acquiring competitive performance with improved efficiency \cite{ref9, ref10}.
The crucial parts of semantic communication include establishing the formal definition of the essential semantics of data and devising the pragmatic extraction methods of compressed semantics from high-dimensional source data.
For the definition of semantics, extensive prior research has been conducted to establish the theory of semantic communication \cite{ref11, ref12, ref13, ref14}.
While each examines subtly distinct concepts, they all share a consistent framework in which source data is encoded into a semantic space at the transmitter side by utilizing the prior knowledge and posterior information of the given data.
Although a fundamental concept of the semantic space had been established in the early works, the specific way to represent the extracted semantics was not adequately defined before the machine learning era \cite{ref9, ref11}.
Built upon the success of machine learning during the past decade, deep learning (DL) has emerged as a key solution for representing data in the compressed embedding space, which serves as a semantic space \cite{ref15, ref16}.
The prior research has mainly focused on semantic communications for transmitting visual data. In the next subsection, we provide details of prior efforts in semantic communications for computer vision (CV). 

\IEEEpubidadjcol
\subsection{Related Works}
\textbf{Semantic Communications for Computer Vision:} 
In the early works of DL-based semantic communication, the joint source-channel coding (JSCC) method was introduced, where it employs convolutional neural networks (CNNs) for extracting a visual semantic, which is the feature vector of a given input image \cite{ref15, ref16}.
The DL-based method demonstrates higher compression rates than traditional image compression techniques while achieving better image reconstruction performance in low signal-to-noise (SNR) and limited bandwidth regimes.
Similarly, the cutting-edge model architecture called Transformer \cite{ref17} is utilized for acquiring rich semantics while surpassing the performance of traditional coding schemes in low SNR conditions \cite{ref15, ref18}.
These approaches hold promise for integrating DL-driven methodologies with semantic communication frameworks \cite{ref9, ref19}, where the extracted feature is transmitted and decoded for reconstructing the original data.

\begin{figure*}[t] 
    \centering
    \includegraphics[width=1.00\textwidth]{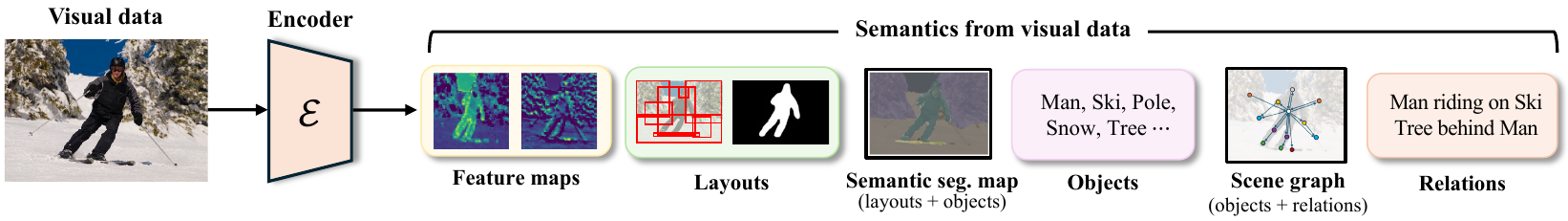}    
    \caption{Semantic features from a given visual data, including feature maps, scene graphs, semantic segmentation maps, layouts, and objects}
    \label{fig:figure1}
\end{figure*}

\textbf{Generative Models for Semantic Communications:}
On the other side, with the substantial progress in deep generative models such as generative adversarial networks (GANs) and diffusion models, numerous researchers have explored their applications in semantic communication, especially for visual data reconstruction at the decoder side \cite{ref20,ref21,ref22}. 
In the cases of \cite{ref23,ref24,ref25,ref26,ref27,ref28}, specific semantic features, such as \textit{feature map}, \textit{semantic segmentation map}, and \textit{scene graph (SG)} are extracted from the source image, then used to reconstruct image through the generative models.
The reconstructed image is then utilized for CV tasks in some studies \cite{ref23,ref24,ref28}. These studies have demonstrated the effectiveness of utilizing semantic features for image transmission and executing CV tasks. 
For instance, the authors of \cite{ref27} employed semantic segmentation maps in conjunction with GANs for image transmission, achieving higher compression rates and improved peak signal-to-noise ratio (PSNR) compared to traditional image compression techniques.
Built upon this approach, the authors of \cite{ref23} considered masked feature maps and introduced reinforcement learning (RL) to dynamically adjust the quantization ratio, thereby further reducing the number of transmitted bits. 
As a different approach, \cite{ref24} divided the semantic segmentation map into multiple one-hot encoded maps and applied conventional image compression to them for efficient communication.
On the other hand, the authors of \cite{ref25} leveraged edge maps and textual information, considering the semantic quality and latency constraints. 
Furthermore, SG-based semantic transmission is explored by employing graph encoding and decoding techniques, outperforming traditional coding schemes in
low SNR conditions \cite{ref26}.
Finally, the authors of \cite{ref29} also utilized SG while introducing the multi-agent RL-based resource allocation algorithm.

\textbf{Multi-Semantic and Multi-Task:}
Recently, some prior works have aimed to utilize multiple semantics, including the aforementioned ones, such as semantic segmentation maps and SG.
Also, researchers have recently started to consider task-dependent transmission, whose purpose is to transmit task-specific adaptation of semantics.
In \cite{ref30}, the authors tackled multi-task scenarios by developing a unified decoder for the multiple semantics. The unified decoder amalgamates the semantic vectors received from the separate encoders, facilitating a cohesive decoding process that efficiently synthesizes information from several source data modalities. 
In \cite{ref23}, the authors modified the quantization levels of transmitted masked feature maps according to particular CV tasks, including classification, detection, and segmentation. 
\subsection{Limitations of Prior Works and Key Motivations}

\textbf{Limitations:} While existing studies in generative model-based semantic communication have shown notable efficiency, they primarily focus on a fixed form of semantics without considering the target tasks simultaneously.
These approaches neglect an important aspect: 
What are the task-adaptive semantics required to run different CV tasks successfully?
The existing methods mostly focus on particular semantics through a semantic encoder and only aim to regenerate the original data through a semantic decoder.
They lack a framework to understand task-adaptive semantics for processing different CV tasks.
In \cite{ref23}, a quantization level of the masked feature map is adjusted for each task without considering the multiple forms of semantics. 
Also, in \cite{ref30}, the authors focused on adjusting the number of transmitted feature maps for different tasks.
To the best of our knowledge, none of the prior works show a thorough examination of how well the CV tasks are being completed for the different choices of visual semantics while considering transmission throughput and task performance.
These limitations increase the communication burden, as extraneous or less relevant semantic information may be transmitted for a given target task.

\textbf{Semantics of Visual Data:} To figure out sufficient information for each task, we illustrate the well-known types of semantic features for a given visual data in Fig. \ref{fig:figure1}.
A list of recognized \textit{objects} is the minimal semantics of the given image, e.g., man, ski, pole, etc.
Also, \textit{layouts} are in the form of bounding boxes or pixel-wise identifications, which provide localization semantics.
When objects and pixel-wise layouts are combined, they form a \textit{semantic segmentation map}, a popular visual semantics used in various CV tasks.
As another semantics, \textit{relations} specifies the particular relations or predicates between the recognized objects, e.g., \texttt{"Man `riding on' Ski"}.
When objects and relations are combined, they construct a \textit{scene graph}, which describes the recognized objects and their relations.
Finally, \textit{feature maps} are the extracted semantic information that bears overall patterns and spatial information of the given image.

\begin{figure}[t] 
    \centering
    \includegraphics[width=\columnwidth]{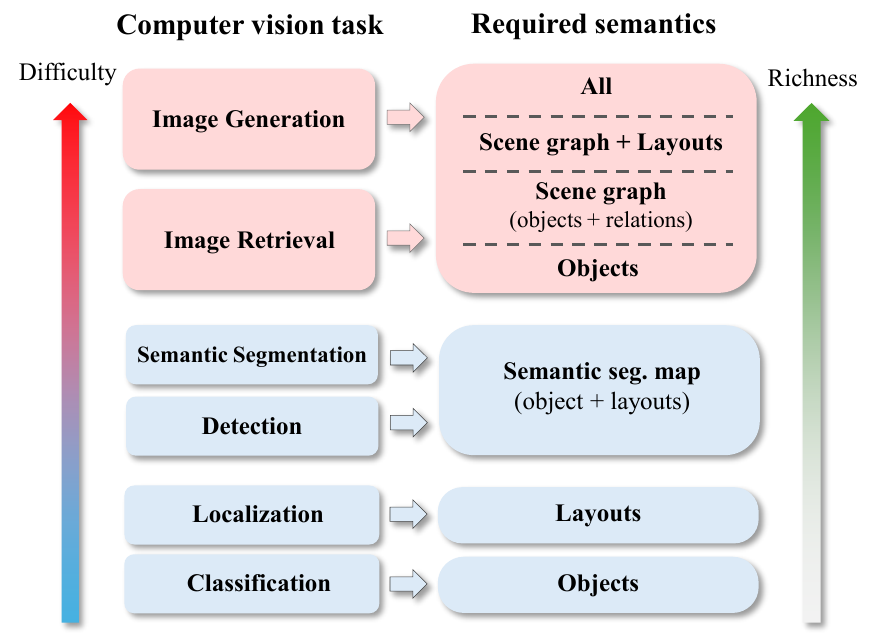}
    \caption{Task-dependent required semantics for computer vision tasks}
    \label{fig:figure2}
\end{figure}
\textbf{Required Semantics for Different CV Tasks:} 
When we complete a CV task at the receiver side, e.g., a receiver desires to localize the objects captured by the camera at the transmitter side, 
a transmitter selects proper semantics and sends them to the receiver, e.g., a list of objects and their layouts, which are sufficient for detection, are transmitted. In this sense, CV tasks can be categorized based on the required semantics.
In Fig. \ref{fig:figure2}, we present the scenarios for transmitting semantics to run different tasks, ranging from classification to image regeneration. Also, we elaborate on the details in Section \ref{sec:prelimB}.

\subsection{Contributions}
Our key contributions include:
\begin{itemize}
\item{We have conducted an extensive study of task-adaptive semantic communications for CV tasks that reveals the required semantic features for the given task for accomplishing acceptable performance.}
\item{We propose a semantic feature filtering algorithm, particularly for scene graphs, which aims to minimize semantic redundancy in the scene. 
This approach maintains the fidelity of the regenerated image while significantly reducing the transmitted data volume.}
\item{We present experimental results demonstrating the effectiveness of the proposed task-adaptive approach in ranging from classification tasks to image generation and retrieval tasks, demonstrating the potential of our approach. Based on the approach, we confirm more than 45 times larger throughput through the efficient selection of visual semantics than a naive transmission of original data in the case of an image generation task.}

\end{itemize}

\subsection{Outlines}
Our paper is organized as follows:
In Section \ref{sec:prelim}, we provide the preliminaries for the further clarification of visual semantics and CV tasks and the basics of diffusion models, which recently work as a semantic decoder. 
In Section \ref{sec:method}, the proposed task-adaptive semantic communication is fully described.
Next, in Section \ref{sec:exp}, we present the extensive simulation results to confirm the efficacy of our framework, including the qualitative and quantitative demonstrations in wireless channel settings.
Section \ref{sec:conclusion} concludes the paper.

\section{Preliminaries}\label{sec:prelim}
\subsection{Formulations of Visual Semantics}\label{sec:prelimA}
Let us consider an image sample $\mathbf{x}\in\mathbb{R}^{M\times N}$, where $M$ and $N$ are the width and the length of $\mathbf{x}$. We then formulate the various visual semantics of the sample as follows:
\begin{itemize}
\item{\textbf{Objects:} These are the individual entities representing tangible items or subjects that can be recognized and classified. As the most fundamental components of an image, objects serve as the building blocks for understanding and interpreting the visual content.
We denote the objects as: $\mathcal{O}(\mathbf{x})=\{o_i\}_{i=1}^{I}$, where $I$ is the number of objects identified through a semantic encoder. Here, $o_i$ can be an object index or text, such as \texttt{"Man"}.
}

\item{\textbf{Relations:}  These semantics capture the interactions and connections between different objects, providing context and understanding of how they relate to one another within the image. These relations often involve structural information that may not be accompanied by a simple layout, providing higher semantic fidelity of the image. Formally, we define the relations as: 
$\mathcal{R}(\mathbf{x})=\{r_{ij}\}$, where $1\leq i,j\leq I$ and $i\neq j$. Here, $r_{ij}$ can be a relation index or textual description, such as \texttt{"riding"}.
Denote $R=|\mathcal{R}(\mathbf{x})|$ as the number of relations.
}

\item{\textbf{Layouts:} These refer to the spatial arrangement of objects within the image, which plays a crucial role in conveying the overall structure and organization of the image. Examples of layouts include bounding boxes, which delineate the position of objects; edge maps, which highlight the boundaries and contours of objects; and segmentation maps, which provide detailed descriptions of object regions.
We formulate layouts as follows:
$\mathcal{L}(\mathbf{x})=\{l_i\}_{i=1}^{I}$, where $l_i=\{(m,n)\in\mathbb{R}^{M\times N}\: |\: (m,n)\mapsto o_i\}$. 
Here, $\mapsto$ indicates that the pixel $(m,n)$ belongs to the object $o_i$.
A layout is a two-dimensional (2D) matrix.
}

\item{\textbf{Semantic Segmentation Maps:} These indicate the identified objects with their spatial locations. For the formula, we denote the semantic segmentation map as $\mathcal{M}(\mathbf{x})=\{(o_i,l_i)\}_{i=1}^{I}$, by simply pairing the objects and layouts.
}

\item{\textbf{Scene Graphs:} These present the objects and their relations in the form of graphs. We formalize a scene graph as $\mathcal{G}(\mathbf{x})=\{(V,E)\}$, with its nodes (or vertices) $V=\{v_i\}$ and edges $E=\{e_{ij}\}$. Here, each node indicates each object, i.e., $v_i=o_i$, and each edge corresponds to each relation, i.e., $e_{ij}=r_{ij}$.
For another form, we can partition the graph into individual sub-graphs, $g_k$, corresponding to a pair of nodes $(o_i,o_j)$ and their relation $r_{ij}$:
\begin{equation}\label{eq:SG}
\mathcal{G}(\mathbf{x}) =\{ g_k = (o_i, r_{ij}, o_j) \mid o_i, o_j \in \mathcal{O}, r_{ij} \in \mathcal{R}\}.
\end{equation}
Each sub-graph can be presented as a sentence, such as \texttt{"Man is riding on ski"}.
}

\item{\textbf{Feature Maps:} These are the representations that entail detailed characteristics and attributes of the image, typically obtained through pretrained deep visual representation models, such as the ResNet architecture \cite{ref31}.
A feature map is a tensor, i.e., $\mathcal{F}(\mathbf{x})\in\mathbb{R}^{C\times X\times Y}$, where $C$ is the channel size, $X$ and $Y$ are the width and the length of the feature map, as respectively. 
}
\end{itemize}
\subsection{Scenarios of Visual Tasks}\label{sec:prelimB}
We describe when each CV task is requested and how each task can be accomplished in the communication systems.
We assume that a transmitter (Tx) captures visual data $\mathbf{x}$ through a sensor system, such as cameras, and a receiver (Rx) requests a sufficient semantic in order to run a given CV task $\mathcal{T}$. Here, we drop $\mathbf{x}$ in the notations of semantics for simplicity.

\begin{figure*}[t] 
    \centering    
    \includegraphics[width=0.95\textwidth]{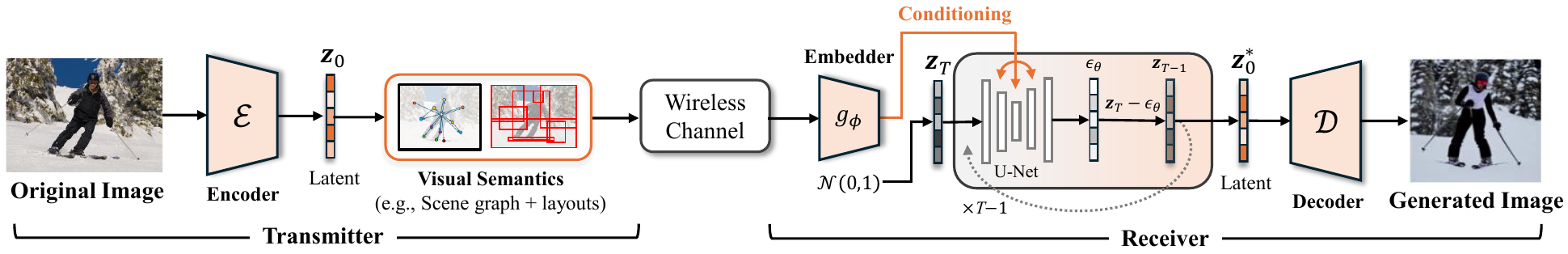}
    \caption{Image generation task with a latent diffusion model by the conditioning from visual semantics (e.g., a scene graph and layouts)}
    \label{fig:figure3}
\end{figure*}

\begin{itemize}

\item{\textbf{Classification:}
The Rx hopes to identify the objects from the visual data. In this case, the Tx is sufficient to transmit the recognized objects, i.e., $\mathcal{O}$, to the Rx.
}

\item{\textbf{Localization:}
The Rx desires to recognize the spatial regions of visual information where some objects are identified; it requires class-agnostic detection. The Tx is then sufficient to transmit the layouts, i.e., $\mathcal{L}$, to the Rx.
As one of the practical scenarios, a vision system that aims to detect the appearance of objects can send the layout to the Rx for localization.
}

\item{\textbf{Detection:}
The Rx wishes to fully detect the objects and their spatial locations. In this case, it is sufficient to transmit the semantic segmentation maps, i.e., $\mathcal{M}$, or the combination of $\mathcal{O}$ and $\mathcal{L}$. As a practical scenario, an alert system using cameras at industrial sites can preserve workers' privacy by sending semantic segmentation maps instead of full images containing their identities.

}
\item{\textbf{Semantic Segmentation}
The Rx expects to identify the objects and their precise spatial locations. In this case, it is sufficient to transmit $\mathcal{M}$, or the combination of $\mathcal{O}$ and detailed $\mathcal{L}$. As a practical scenario, in autonomous driving, the system can detect pedestrians or vehicles by transmitting the semantic segmentation maps to the Rx, ensuring safety without sending full image data.}

\item{\textbf{Image Retrieval:}
The Rx pursues picking out or retrieving images with semantics similar to the visual information captured on the Tx side.
It can be done with simple objects $\mathcal{O}$, but additional layouts $\mathcal{L}$ possibly add spatial information to it (equivalent to sending semantic seg. maps $\mathcal{M}$), and relations $\mathcal{R}$ makes the retrieved images more similar to the original one in presenting the relations between objects (equivalent to sending scene graph $\mathcal{G}$). Finally, the feature map $\mathcal{F}$ provides the full semantic information of the image. As a practical scenario for virtual reality, many portions of visual information do not have to be exactly the same as the original one, but it is sufficient to keep the semantic information the same. In this case, the Tx detours the full image transmission but sends sufficient semantics to run image retrieval.
}

\item{\textbf{Image Generation:}
The Rx aims to regenerate data as identical as possible to the original visual data at the Tx.
To this end, objects $\mathcal{O}$, layouts $\mathcal{L}$, relations $\mathcal{R}$, and the feature map $\mathcal{F}$ may need to be transmitted to the Rx. Richer semantics would be beneficial to accurate regeneration via semantic decoders.
The highly accurate replication of visual information, such as the Metaverse, can be a practical scenario for image generation.
}

\end{itemize}

\subsection{Generative Models in Semantic Communications}
In semantic communication systems, generative models are widely utilized to transform compressed semantic features into high-quality visual images, enabling the transmission of only the essential semantic information. 
In this sense, the generative model itself is often considered as the semantic decoder, serving as a part of the communication system designed to execute tasks \cite{ref23, ref24, ref28}.
We present here the basics of generative models, particularly in diffusion models, as the semantic decoder.
While various kinds of generative models have been explored in the research field, the diffusion model, known for its outstanding image generation capabilities, has been widely adopted in recent research \cite{ref24, ref25, ref26}.
In diffusion models, clean image data is generated from a noisy image by gradually removing the additive noise. 
Thus, the diffusion process is described by the forward process and the reverse process. In the forward process, the original image \(\textbf{x}_0\) is corrupted by Gaussian noise at each time step \(0<t \leq T\)  until it becomes pure noise \(\textbf{x}_T\). The noised image \(\textbf{x}_t\) at time step \(t\) is defined as:
\begin{equation} 
q(\textbf{x}_t \mid \textbf{x}_{t-1}) = \mathcal{N}(\textbf{x}_t; \sqrt{1 - \beta_t} \textbf{x}_{t-1}, \beta_t \mathbf{I}),
\end{equation}
where $\mathcal{N}(\:\cdot\:;\mu,\sigma^2)$ is the Gaussian distribution of mean $\mu$ and variance $\sigma^2$.
In the reverse process, starting from \(\textbf{x}_T\), the model parameterized with \(\theta\) to estimate the error, i.e., $\epsilon_{\theta}$, iteratively denoises the image by predicting the added noise at each time step\cite{ref22}. 
The loss is defined to minimize the difference between the true noise \(\epsilon\) and the predicted noise \(\epsilon_{\theta}\):
\begin{equation} L := \mathbb{E}_{t, \textbf{x}_0, \epsilon} \left[ |\epsilon - \epsilon_{\theta}(\textbf{x}_t, t)|^2 \right].
\end{equation}
For a more efficient diffusion process, the latent diffusion model (LDM) was proposed \cite{ref31}.
Instead of applying the diffusion process directly in pixel space, LDM performs diffusion in a lower-dimensional latent space learned by a pretrained  VAE.
By using a latent representation of the data, i.e., $\mathbf{z}_t$, the model can focus on the essential semantics of the image, allowing faster and more efficient generation without sacrificing quality. 
In addition to its efficiency, it is capable of generating images conditioned on a variety of semantic information.
The loss function of LDM-based conditioning is expressed as:
\begin{equation}
\label{eq3}
L_{\text{LDM}} := \mathbb{E}_{t, \mathbf{y}, \epsilon} \left[ \|\epsilon - \epsilon_\theta(\mathbf{z}_t, t, g_{\phi}(\mathbf{y}))\|^2 \right],
\end{equation}
where \(\mathbf{y}\) corresponds to the conditions and $g_{\phi}$ embeds the conditions to the latent space.
In Fig. \ref{fig:figure3}, we further envision how the conditioned generation of LDM can be used in semantic communications, where the received semantics are used as conditioning.
When the Tx transmits a scene graph, i.e., $\mathcal{G}$, and layouts, i.e., $\mathcal{L}$, the Rx utilizes the semantics as the conditioning to be injected into the diffusion process through the attention layers of the U-Net, guiding the denoising process to ensure the generated image aligns with the given condition. 

\begin{figure*}[t] 
    \centering
    \includegraphics[width=0.9\textwidth]{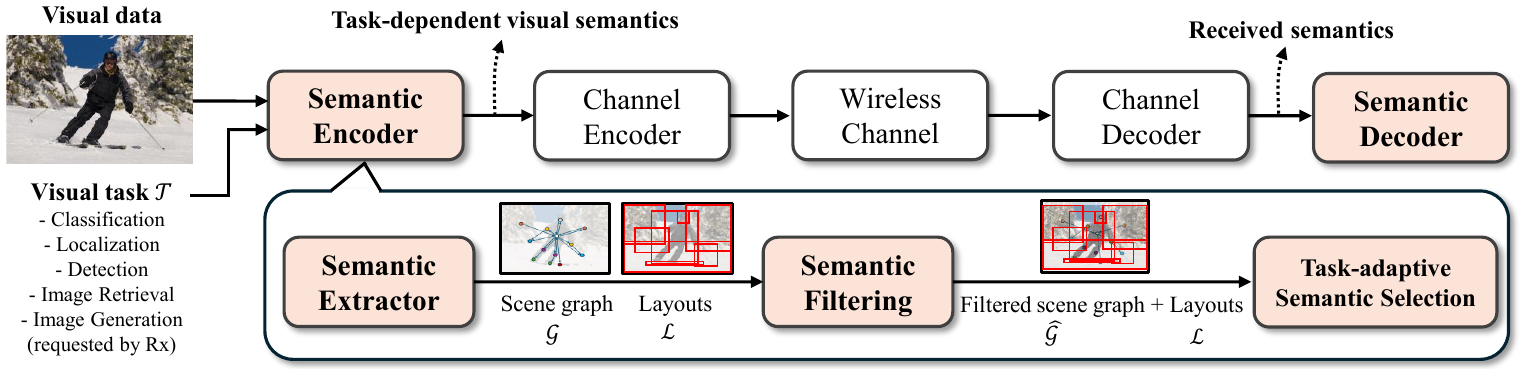}    \caption{The proposed framework of the semantic encoder part. Here, we illustrate the case of transmitting the scene graphs, i.e., $\mathcal{G}$ (objects + relations), and the layouts, i.e., $\mathcal{L}$. After filtering out redundant sub-graphs to obtain $\mathcal{G}_{\text{filtered}}$, the encoder sends the selected semantics through the wireless channel.}
    \label{fig:figure4}
\end{figure*}
\section{Proposed Task-Adaptive Semantic Communication Systems}\label{sec:method}
In this section, we provide a full description of our proposed semantic communication systems for CV tasks.
As aforementioned in Section \ref{sec:prelim}, our communication scenario is initiated as follows: For a given pair of a transmitter (Tx) and a receiver (Rx), the Rx aims to complete a certain CV task for the given visual data at the Tx side, and it requests the semantic information of the data.

First, the Tx begins to process to obtain the proper visual semantics of the data, which is tailored to complete the requested CV task by the Rx. Specifically, Fig. \ref{fig:figure4} illustrates the process, particularly for the \textit{semantic encoder} part of the Tx.
Semantic Encoder consists of three main blocks: \textbf{i)} Semantic extractor block, \textbf{ii)} Semantic filtering block, and \textbf{iii)} Task-adaptive semantic selection block. 
The semantic extractor block is for extracting semantic features from the input image. 
As listed in part \ref{sec:prelimA}, the semantic extractor is capable of obtaining objects, i.e., $\mathcal{O}$, layouts, i.e., $\mathcal{L}$, relations, i.e., $\mathcal{R}$, semantic segmentation maps (objects + layouts), i.e., $\mathcal{M}$, scene graphs (objects + relations), i.e., $\mathcal{G}$, and feature maps, i.e., $\mathcal{F}$.
The semantic filtering block filters the semantics for dropping less informative or redundant information. 
In the extracted semantics, some are trivial, so thus less informative, or some are inferable from others, so thus redundant. Here, we focus on filtering a scene graph, which yields a significant gain in transmission efficiency without losing utility.
Finally, the task-adaptive semantic selection block ensures that only the semantic features required for the CV task are transmitted. The selection process is guided by the specific requirements of the target task, and in some cases, task performance considerations are also factored into selection. 
We further provide detailed explanations of each block, highlighting its roles and contributions to our framework.

\subsection{Semantic Extractor}
For a given image $\mathbf{x}$, the semantic extractor block begins with obtaining the feature map, i.e., $\mathcal{F}(\mathbf{x})$ through a feature extractor architecture. 
In our system, we utilize the ResNet architecture as the feature extractor for the experiments, but it is not limited to using the particular model architecture \cite{ref32}.

From the feature map, as mostly done by vision models, the following semantics are computed: Objects $\mathcal{O}(\mathbf{x})$, Layouts $\mathcal{L}(\mathbf{x})$, and Relations $\mathcal{R}(\mathbf{x})$. In our framework, we use the Transformer architecture to acquire the aforementioned semantics, which is the cutting-edge method for computing visual semantics. Also, layouts can be the forms of bounding boxes or pixel-wise layouts. In our experiments, we utilize different layout forms for varying tasks.

Based on the obtained semantics, the semantic segmentation maps, i.e., $\mathcal{M}(\mathbf{x})$, can be computed by simply using the objects and the pixel-wise layouts.

In addition, the scene graph, i.e., $\mathcal{G}(\mathbf{x})$, is computed by combining the objects and the relations by following the scene graph generation process in \cite{ref33}.
Each node is assigned to each object, and each edge corresponds to each relation, as formulated in Section \ref{sec:prelimA}. For further processing, the graph can be transformed into a set of sub-graphs by following Eq. \eqref{eq:SG}.
By partitioning $\mathcal{G}(\mathbf{x})$ into an individual sub-graph, i.e., $g_k$, each sub-graph can be represented with a sentence about two objects, $o_i$, $o_j$, and their relation $r_{ij}$. 

\subsection{Proposed Semantic Filtering and Its Theory}
Before presenting our semantic filtering algorithm, we first introduce the underlying theory behind it. For clarity, consider a simple communication scenario: a Tx aims to convey its observations to an Rx. Here, the observation of Tx can be represented as a set of statements $W_s = \left\{ w_k \right\}_{k=1}^{K}$ such that $W_s\subseteq W$, where $W = \left\{ w_k \right\}_{k=1}^{Z}$ is the all statements or semantics that Tx can potentially observe from a given input. 
For example, $w_k$ can be an extracted semantic from an image, e.g., \texttt{"Man is riding on ski"} in a scene graph. In this context, a sub-graph $g_k$ from the semantic extractor can be understood as a statement $w_k$ in the given image.

Intuitively, an efficient Tx would not merely send all $w_k$. Instead, it would filter out less informative or semantically redundant statements from $W_s$; thus, obtaining $\hat{W}_s$, where $\hat{W}_s \subseteq W_s$. 
The objective underlying the filtering is to maximize communication efficiency, which reconciles enlarging mutual information $I(\hat{W}_s;{W}_s)$, and reducing the size of $\hat{W}_s$:
\begin{equation}\label{eq:opt_eff}
\max\limits_{\hat{W}_s\subseteq W_s}{\frac{I(W_s;\hat{W}_s)}{c \cdot |\hat{W}_s|}},
\end{equation}
Here, we assume that the communication cost (in bits) for transmitting each statement is $c$, and $I(\cdot;\cdot)$ denotes the mutual information of two events.
When considering a deterministic filtering function denoted as $q(\hat{W}_s | W_s)$, we know that $H(\hat{W}_s|W_s)=0$; thus,
$I(W_s;\hat{W}_s)=H(\hat{W}_s)-H(\hat{W}_s|W_s)=H(\hat{W}_s)$.
Based on it, we further formulate Eq. \eqref{eq:opt_eff} as follows:
\begin{gather}
\max_{q(\hat{W}_s\mid W_s)} \; \mathop{\mathbb{E}}_{\mathbb{P}(W_s)}\mathop{\mathbb{E}}_{q(\hat{W}_s \mid W_s)} \left[ \frac{\log\left( \frac{\mathbb{P}(W_s,\hat{W}_s)}{\mathbb{P}(W_s)\,\mathbb{P}(\hat{W}_s)} \right)}{c \cdot |\hat{W}_s|} \right]\\
= \max_{q(\hat{W}_s \mid W_s)} \; \mathop{\mathbb{E}}_{\mathbb{P}(W_s)}\mathop{\mathbb{E}}_{q(\hat{W}_s \mid W_s)} \left[ \frac{-\log \mathbb{P}(\hat{W}_s)}{c \cdot |\hat{W}_s|} \right] \label{eq:7} \\
= \max_{q(\hat{W}_s \mid W_s)} \; \mathop{\mathbb{E}}_{\mathbb{P}(W_s)}\mathop{\mathbb{E}}_{q(\hat{W}_s \mid W_s)} \left[ \frac{-\log \mathbb{P}(W_s)q(\hat{W}_s \mid W_s)}{c \cdot |\hat{W}_s|} \right].
\end{gather}
Based on the formulation, the optimal $q^*(\hat{W}_s | W_s)$ can be found by evaluating the efficiency values for every possible $\hat{W}_s$, but it is infeasible to check all possible $2^K-1$ cases.

Let us simplify the problem to focus on deciding whether to filter out a single statement from $W_s$ or not.
When Tx filters out a single $w_K$ from $W_s$ (without loss of generality, we pick the statement of the last index $K$). 
\[
q\bigl(\hat W_s \mid W_s\bigr)
=
\begin{cases}
1, & \hat W_s = \hat{W_s^*} := \{w_k\}_{k=1}^{K-1},\\
0, & \text{otherwise}.
\end{cases}
\]
If the filtering is effective, it must improve the efficiency (by focusing on the term in Eq. \eqref{eq:7}):
\begin{equation}
 \frac{-\log \mathbb{P}(\hat W^*_s)}{c(\left|W_s\right|-1)} >\frac{-\log \mathbb{P}({W}_s)}{c\left|{W}_s\right|}
\end{equation}
\if false
\begin{equation}
{\left|{W}_s\right|\log \mathbb{P}(\hat{W}^*_s)} <{{\left(\left|{W}_s\right|-1\right)}\log \mathbb{P}({W}_s)}
\end{equation}
\fi
\begin{equation}
{\log\left( \mathbb{P}(\hat{W}^*_s)^{\left|{W}_s\right|}\right)} <{\log\left( \mathbb{P}({W}_s)^{\left|{W}_s\right|-1}\right)}.
\end{equation}
We can use chain rule to express $\mathbb{P}(\hat{W}_s)$ and $\mathbb{P}({W_s})$: 
\begin{align}
\mathbb{P}({w}_1)^{\left|{W}_s\right|}...&\mathbb{P}({w}_{K-1}|w_{[K-2]})^{\left|{W}_s\right|}\nonumber\\
&<\mathbb{P}({w}_1)^{\left|{W}_s\right|-1}...\mathbb{P}({w}_K|w_{[K-1]})^{\left|{W}_s\right|-1}
\end{align}
\begin{equation}
\log{\mathbb{P}(\hat{W}^*_s)}<\left(\left|{W}_s\right|-1\right)\log{\mathbb{P}({w}_K|w_{[K-1]})}
\end{equation}
\begin{equation}\label{eq:imp_eff}
\frac{\log{\mathbb{P}(\hat{W}^*_s)}}{\left|{W}_s\right|-1} < \log{\mathbb{P}({w}_K|\hat{W}_s^*)},
\end{equation}
where $[K]$ indicates a set of integers less than or equal to $K$, and $w_{[K]}=\{w_1,\cdots,w_{K}\}$.
Eq. \eqref{eq:imp_eff} states that if it holds, Tx can filter out $w_K$, which can be easily inferred from other statements, while reducing the communication cost. 
With this understanding, Tx can iteratively filter out statements until the inequality is nearly collapsed. 
In practice, however, the $\mathbb{P}(\hat{W}^*_s)$ is intractable to compute. 
To address this, we adopt a simple yet effective algorithm that can be implemented by using the conditional probability and an arbitrary threshold $\tau$: filtering out $w_K$, if $\mathbb{P}(w_K|\hat{W}_s^*)>\tau$.
Intuitively, we aim to filter out a certain statement if it can be inferred from others with a sufficiently high probability, while considering the cost.

We attempt to fundamentally analyze the semantic filtering strategies by considering the extracted features as a form of statements. For feature maps, which are widely used in prior works \cite{ref15, ref16, ref18, ref23}, it is infeasible to analyze the redundancy of each component feature. 
In contrast, we utilize the conditional probability, $\mathbb{P}(w_K|\hat{W}_s^*)$, which directly implies the redundancy of each component statement.

Let us then specify our semantic filtering algorithm. We particularly focus on the scene graph $\mathcal{G}$ as semantics, since other semantics show their own unique information that is disjoint from others.
For example, the layouts are the only semantics that bear spatial information. Likewise, objects and relations present the categories of recognized objects and their relations, which contain sufficiently compact information. 
On the other hand, the scene graph $\mathcal{G}$, which merges objects and relations, may contain less informative or redundant subgraphs that can be dropped in transmission. For other semantics, such as feature maps and semantic segmentation maps, prior methods exist to mask them to compress the information.

\begin{figure*}[t] 
    \centering
    \includegraphics[width=1\textwidth]{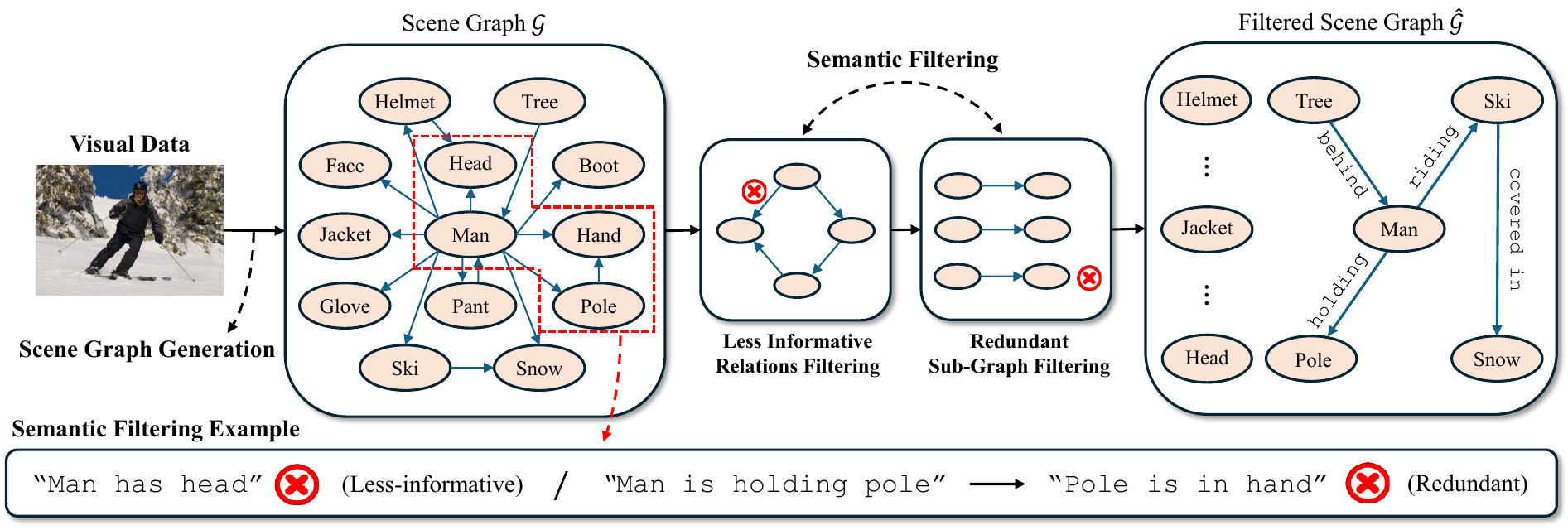}    
    \caption{Semantic Filtering}
    \label{fig:sg_filtering}
\end{figure*}

Consequently, we set the statements to be the form of scene graph: $W_s = \mathcal{G} = \left\{g_k \right\}_{i=1}^{K}$, where the probability of scene graph $\mathcal{G}$ is formulated as follows.
\begin{equation}
\mathbb{P}(\mathcal{G}) = \mathbb{P}(g_1,g_2,\dotsc,g_K) = \prod_{k=1}^{K} \mathbb{P}(g_k \mid g_1, \dotsc, g_{k-1})  
\end{equation}
To further specify the proposed semantic filter algorithm for scene graphs, we consider two aspects of filtering: less informative relation filtering and redundant sub-graph filtering. Fig. \ref{fig:sg_filtering} illustrates the steps to obtain the filtered scene graph.

\textbf{Less Informative Relations Filtering:} Assuming that the probability of observing each object in an image is independent, the probability of sentence or sub-graph \(g_k\) is:
\begin{equation}
\mathbb{P}(g_k) = \mathbb{P}(o_i, r_{ij}, o_j) = \mathbb{P}(r_{ij}\:|\:o_i, o_j) \cdot \mathbb{P}(o_i) \cdot \mathbb{P}(o_j).
\end{equation}
Considering that each element compromising $g_k$ is also seen as an individual statement, we can drop the relation (edge of graph), which can be commonly expected under the observance of two objects: \(\mathbb{P}(r_{ij}\:|\:o_i, o_j) \approx 1\).
Note that it does not represent the marginal probability, i.e., $\mathbb{P}(r_{ij})$, but the conditional probability of $r_{ij}$, given two objects $o_i$ and $o_j$.
For instance, when a \texttt{"man has head"} is detected in the image, the conditional probability of the relation \texttt{"has"} approximates to 1, indicating that the relation carries low information. 
On the contrary, let us consider another example of \texttt{"cat wearing hat"}. If the sender has rarely observed a cat wearing a hat, the conditional probability of the relation \texttt{"wearing"} would be low, suggesting that this observation carries a high amount of information, regardless of how explicit or obvious the relationship appears.
Thus, the semantic filtering block drops $r_{ij}$ with sufficiently large conditional probability\footnote{In experiments, we compute it based on the frequency to appear $r_{ij}$ given $o_i$ and $o_j$, in a given dataset or knowledge base.}, as described in Algorithm \ref{alg1}.

\begin{algorithm}[t]
\caption{Less Informative Relations Filtering Algorithm}
\begin{algorithmic}[1]
\REQUIRE A scene graph \(\mathcal{G}=\{g_k\}\), threshold $0\leq\tau_{f}\leq 1$
\ENSURE A filtered scene graph $\hat{\mathcal{G}}$
\STATE Initialize $\hat{\mathcal{G}} \leftarrow {\mathcal{G}}$
\FOR{each sentence $g_k = (o_i, r_{ij}, o_j) \in \mathcal{G}$}
    \STATE Compute $\mathbb{P}(r_{ij}\:|\:o_i, o_j)$
    \IF{$\mathbb{P}(r_{ij} \:|\: o_i, o_j) \geq \tau_{f}$}
        \STATE Remove $r_{ij}$ from $\hat{\mathcal{G}}$
    \ENDIF
\ENDFOR
\STATE \textbf{return} $\hat{\mathcal{G}}$
\end{algorithmic}\label{alg1}
\end{algorithm}

\textbf{Redundant Sub-Graph Filtering:} 
Next, we can further compress scene graphs by filtering out redundant sub-graphs.
When a sub-graph or sentence $g_k$ is trivially inferred with given other sub-graphs, i.e., \( \mathbb{P}(g_k\:|\: g_1, \dotsc, g_{k-1}) \approx 1\), then $g_k$ does not need to be transmitted.
Let us go back to Fig. \ref{fig:sg_filtering}. We can trivially infer a sub-graph or sentence \texttt{"Pole is in hand"} given \texttt{"Man is holding pole"}.
It shows a sub-graph provides sufficient information to infer another sub-graph.
For this case, we can filter out \texttt{"Pole is in hand"} by selecting \texttt{"Man is holding pole"}.
In another case, multiple sub-graphs can be used together to infer another one. 
For example, with the combination of \texttt{"man is riding horse"} and \texttt{'tree is behind man'}, we can trivially infer \texttt{"tree is behind horse"}.

However, there remains a critical challenge in filtering out inferable sub-graphs. While frequency‐based estimation of conditional probabilities works well for simple relation filtering, the combinatorial explosion of sub‐graph combinations in a full scene graph renders counting‐based estimation computationally intractable. In the aforementioned cases, humans easily figure out the redundant one based on common sense: Man often holds something with their hands, and the horse with man is behind the tree because the man is behind it.

To utilize the in-depth semantics behind each sub-graph, we leverage a pretrained language model, such as Sentence-BERT (S-BERT) in \cite{ref34}, to recognize redundant sentences.
As with other recent language models, the key characteristic of S-BERT is its ability to encode sentences or phrases into dense embedding vectors so that the relationships between them can be measured effectively. 
In particular, when two embedding vectors represent semantically entailing sentences, the cosine similarity or Euclidean distance between them becomes minimal. 
On the other hand, if the sentences are in a contradictory relationship, these metrics yield larger distances. 
In cases where the relation is neutral, the vectors are close to being perpendicular in the vector space. 
Based on this property, the conditional probability of a sentence $g_k$ based on others can be well-estimated.

When embedding our sub-graph sentence $g_k$ via a language model $F_{\text{LM}}(\cdot)$, the normalized embedded vector is denoted as $\mathbf{g}_k:=F_{\text{LM}}(g_k)/\|F_{\text{LM}}(g_k)\|$.
Let us then consider a case with three vectors $\mathbf{g}_1$, $\mathbf{g}_2$, $\mathbf{g}_3$ from $g_1$, $g_2$, $g_3$, respectively.
When subtracting the direction of $\mathbf{g}_3$ from the other two via projection, the following are obtained: 
\begin{gather}
\mathbf{g}_{1|3} = \mathbf{g}_1 - \text{Proj}_{\mathbf{g}_3}(\mathbf{g}_1) \\
\mathbf{g}_{2|3} = \mathbf{g}_2 - \text{Proj}_{\mathbf{g}_3}(\mathbf{g}_2),
\end{gather}
where $\text{Proj}_{\mathbf{u}}(\mathbf{v})$ is the projected $\mathbf{v}$ on $\mathbf{u}$.
The process removes the information of $\mathbf{g}_3$ from the other two vectors.
From the view of $\mathbf{g}_1$, when further subtracting the information of $\mathbf{g}_2$, we can solely focus on the residual vector, i.e., $\mathbf{r}_1$, of $\mathbf{g}_1$:
\begin{align}
\mathbf{r}_1 := \mathbf{g}_{1|2,3} &= \mathbf{g}_{1|3} - \text{Proj}_{\mathbf{g}_{2}}(\mathbf{g}_{1|3}) \\ &= \mathbf{g}_{1|3} - \text{Proj}_{\mathbf{g}_{2|3}}(\mathbf{g}_{1|3}).
\end{align}
When the Euclidean norm, i.e., $\|\mathbf{g}_{1|2,3}\|$ gets close to zero, 
 \(g_2\), \(g_3\) or combination of them entails the meaning of \(g_1\), which can be interpreted as \(g_1\) can be inferred from other sentences i.e., \( \mathbb{P}(g_1 \:|\: g_2, g_3) \approx 1\). 
 In other words, \(g_1\) is semantically redundant. 
 As a general form of residual vector:
 $\mathbf{r}_{k} = \mathbf{g}_{k|[K]\setminus k}$,
 where $[K]$ indicates a set of integers less than or equal to $K$.
 Algorithm \ref{alg2} presents the generalized form of this approach to identify and filter out semantically redundant features systematically. 
 We denote the set of normalized embedded vectors of sentences $\{g_k\}$ as follows:
\begin{equation}
\mathcal{G}_{\text{LM}} = \left\{ \mathbf{g}_k = \frac{F_{\text{LM}}(g_k)}{\|F_{\text{LM}}(g_k)\|} \mid g_k \in \mathcal{G} \right\}.
\end{equation} 
\begin{algorithm}[t]
\caption{Redundant Sub-Graph Filtering Algorithm}
\begin{algorithmic}[1]
\REQUIRE A set of vectors $\mathcal{G}_{\text{LM}} = \{ \mathbf{g}_k \}$, threshold $0\leq\tau_{r}\leq 1$
\ENSURE Filtered scene graph $\hat{\mathcal{G}}$
\WHILE{True}
    \FOR{each $\mathbf{g}_k \in \mathcal{G}_{\text{LM}}$}
        \FOR{each $\mathbf{g}_j \in\mathcal{G}_{\text{LM}}$}
            \IF{$j \neq k$}
                \STATE Remove information from vector :
                
                \(
                \mathbf{g}_{k|[j]\setminus k} \gets \mathbf{g}_{k|[j-1]\setminus k} - \text{Proj}_{\mathbf{g}_j}(\mathbf{g}_{k|[j-1]\setminus k})
                \)
            \ENDIF
        \ENDFOR
        \STATE Compute the Euclidean norm of the  residual vector:
        
        \(
        \|\mathbf{r}_{k}\| \gets  \|\mathbf{g}_{k|[K]\setminus k }\|
        \)
    \ENDFOR
    \STATE Find $\mathbf{r}_{m}$ with the smallest Euclidean norm.
    \IF{\(
        \| \mathbf{r}_{m}\| 
        \) $< \tau_r$ }
        \STATE Remove $\mathbf{g}_m$ from $\mathcal{G}_{\text{LM}}$
    \ELSE
        \STATE \textbf{break}
    \ENDIF
\ENDWHILE
\STATE $\hat{\mathcal{G}}\leftarrow\{g_k \:|\: \mathbf{g}_k\in\mathcal{G}_{\text{LM}}\}$
\STATE \textbf{return} $\hat{\mathcal{G}}$
\end{algorithmic}\label{alg2}
\end{algorithm}

Even when Tx obtains highly efficient $\hat{W}_s$, e.g., $\hat{\mathcal G}$, filtering can lead to some level of semantic ambiguity $\epsilon$, which we define as $\epsilon \triangleq 1 - \mathbb{P}_{\mathcal{K}, \mathcal{I}}(W_s|\hat{W}_s)$, where $\mathcal{K}$ and $\mathcal{I}$ denote the background knowledge and the inference mechanisms, respectively.
Specifically, $\mathcal{K}$ and $\mathcal{I}$ broadly include the pretrained semantic extractor, the language model used in the sub-graph filtering, and the statistics of statements used in the relation filtering.
In detail, our Rx employs the pretrained generative models as its $\mathcal{K}$ and $\mathcal{I}$.
Importantly, semantic ambiguity perceived by transmitter $\epsilon_{\mathrm{Tx}}$, and the receiver, $\epsilon_{\mathrm{Rx}}$, can differ significantly, as they may possess different $\mathcal{K}$ and $\mathcal{I}$.
For some cases, Rx might trivially reconstruct filtered statements despite the ambiguity from Tx, i.e., $\epsilon_{\mathrm{Tx}} > \epsilon_{\mathrm{Rx}}$.
Conversely, Rx might fail to infer them even when Tx perceives minimal ambiguity, i.e., $\epsilon_{\mathrm{Rx}} > \epsilon_{\mathrm{Tx}}$.
In this sense, the true objective of Tx can be seen as finding $\hat{W}_s$ that minimizes $\epsilon_{\mathrm{Rx}}$ i.e., maximizing the likelihood $\mathbb{P}_{\mathcal{K}_{\mathrm{Rx}}, \mathcal{I}_{\mathrm{Rx}}}(W_s|\hat{W}_s)$ while reducing $|\hat{W}_s|$. However, it is generally unrealistic to assume that Tx has full access to $\mathcal{K}_{\mathrm{Rx}}$ and $\mathcal{I}_{\mathrm{Rx}}$. Our configuration employs a knowledge-rich generative model as the semantic decoder, which aligns with the scenario where $\epsilon_{\mathrm{Rx}} < \epsilon_{\mathrm{Tx}}$. Under this, the upper bound of the semantic ambiguity of our communication system is $\epsilon_{\mathrm{Tx}}$.

It is also worth emphasizing that our filtering approach differs from previous works in two key perspectives:
\begin{itemize}
\item{\textbf{Filtering Perspective:} While previous studies primarily focused on reducing the cost $c$ for a given a set $W_s$ based on marginal probability (not using conditional probability) or devising schemes such as quantizing or resizing the extracted features, e.g., feature maps, segmentation maps, etc \cite{ref23, ref24, ref25, ref27}. Our approach aims to find an efficient $\hat{W}_s$ for a given $W_s$ at \textit{semantic level}, e.g., scene graphs, by utilizing conditional probability $\mathbb{P}(w_K|\hat{W}_s^*)$.}
\item{\textbf{Error Pespective:}
Due to the direct consideration of the redundancy for each  $w_k$, our approach suggests that proper filtering does not necessarily induce errors in semantics, as it drops semantically redundant ones. 
}
\end{itemize}

\subsection{Task-Adaptive Semantic Selection}

Our approach identifies the semantic features needed for each task and transmits them accordingly. The required semantic features of each task are illustrated in Fig. \ref{fig:figure2}. Also, we fully describe the scenarios of each visual task in Section \ref{sec:prelimB} to elaborate on the required semantics for each task.

For simple tasks, such as classification and localization, selecting the required semantics is quite straightforward. More importantly, however, the required semantics for complex tasks such as image generation may vary with the underlying users' intention and the target performance. For example, the Rx may desire to reconstruct the image as accurately as possible, whereas in other scenarios, high-level semantic interpretation of the image content may be sufficient. Transmitting richer semantics, however, increases communication overhead, creating an inherent trade-off between communication efficiency and task performance. 
To our knowledge, no prior work has systematically quantified the costs and benefits of different semantic types when handling complicated vision tasks.
Thus, we conduct extensive evaluations to scrutinize which semantic types should be chosen to transmit within the finite time interval and acquire the level of performance.

Based on the understanding, our system provides a set of visual semantics, including objects $\mathcal{O}( \mathbf{x})$, layouts $\mathcal{L}(\mathbf{x})$, semantic segmentation maps $\mathcal{M}(\mathbf{x})$, filtered scene graphs $\mathcal{G}_{\text{filtered}}(\mathbf{x})$, and feature map $\mathcal{F}( \mathbf{x})$ enabling task-adaptive selection of them.

\section{Experimental Results}\label{sec:exp}
We focus on the following three perspectives of evaluation.
\begin{itemize}
\item{\textbf{Task Performance:} We assess the impact of selective semantic transmission on the most challenging tasks, e.g., image generation and retrieval, focusing on how the selection of different semantic features impacts the results. Furthermore, we confirm the critical importance of transmitting only the required semantics for simpler tasks, e.g., detection and semantic segmentation.}
\item{\textbf{Transmission Efficiency:}
We analyze transmission throughput for each semantic type and quantify the bit-rate savings afforded by our filtering method.}
\item{\textbf{Latency:} We analyze the end-to-end latency for the overall process to verify real-time working. Finally, by benchmarking against existing approaches, we demonstrate the significance of task-adaptive system design in enabling real-time computer-vision tasks.
}
\end{itemize}

\if false
The experiments of our framework are threefold: \textbf{i)} evaluation of task performance to show how much a given visual task is performed well, particularly for the most challenging cases, i.e., image generation and retrieval, \textbf{ii)} demonstration on wireless communication channels to confirm how fast processing of a given visual task is possible, \textbf{iii) analysis on end-to-end latency for the overall process to verify real-time working.
Specifically, we first analyze the task performance across various scenarios where different tasks are required, focusing on how the selection of different semantic features impacts the results. 
We then evaluate the communication efficiency achieved through the selective and efficient transmission of semantic features on wireless channels. 
Lastly, we analyze end-to-end latency for varying tasks to evaluate the real-time working availability of our proposed framework.
\fi
\subsection{Experimental Settings}

\textbf{Datasets:} In the experiments, five different visual datasets are used: the Visual Genome dataset, Open Image V6 dataset, ImageNet-1K dataset, COCO dataset, and Cityscapes dataset \cite{ref35, ref36, ref37, ref38, ref39}. These diverse datasets allow us to comprehensively evaluate our method across a range of images and scenarios. A brief description and their usages are as follows.
 \begin{itemize}
\item{\textbf{Visual Genome Dataset} \cite{ref35}: This dataset was used for image generation and image retrieval tasks. It consists of images annotated with detailed information, such as objects and relationships, making it well-suited for assessing the effectiveness of our system in generating or retrieving images based on semantic features.}
\item{\textbf{Open Image V6 Dataset} \cite{ref36}: This dataset contains diverse scenes, and we use the subset of it for the image generation task to evaluate our semantic communication framework across various contexts.}
\item{\textbf{ImageNet-1K Dataset} \cite{ref37}: This dataset is characterized by simpler contexts and less diverse scenes, and we use the subset of it for the image generation task to evaluate our framework under less complex scenarios.}
\item{\textbf{COCO Dataset} \cite{ref38}: For image classification, localization, and detection, we use the subset of the COCO dataset. It is widely used for various CV tasks and serves as a popular standard benchmark.}
\item{\textbf{Cityscapes Dataset} \cite{ref39}: We use the test split of this dataset in the image semantic segmentation task. This dataset features high-resolution images capturing a diverse range of street scenes, each accompanied by detailed and fine-grained pixel-level annotations.}
\end{itemize}

\textbf{Model Architectures:}
For the semantic extractor, we employ the ResNet-50 encoder \cite{ref32} that transforms a given image $\mathbf{x}\in\mathbb{R}^{3\times M\times N}$ to the feature map, i.e., $\mathcal{F}(\mathbf{x})\in\mathbb{R}^{C\times X\times Y}$. The encoder is pretrained on the ImageNet-1K classification task \cite{ref37}. In our configuration, $M=N=512$, and $X=Y=16$. The channel dimension $C$ is $2048$.
To obtain semantic features described in Section \ref{sec:prelim}, the overall process is as follows: First, we utilize a transformer encoder-decoder architecture to obtain $\mathcal{O}(\mathbf{x})$ and $\mathcal{L}(\mathbf{x})$ from $\mathcal{F}(\mathbf{x})$. However, as $\mathcal{F}(\mathbf{x})$ it has a high-dimensional channel space, which is less suitable for transformer architecture, 1$\times$1 convolution is applied  $\mathcal{F}(\mathbf{x})$ to reduce channel dimensions $C$ to a smaller $d=256$. Then, the feature map is flattened resulting $\mathcal{F}_0(\mathbf{x})\in\mathbb{R}^{d\times XY}$. To encode spatial information, we add a fixed positional embedding $E_p\in\mathbb{R}^{d\times XY}$ to the feature map, resulting in $\mathcal{F}_p(\mathbf{x})$. The encoder part utilizes a self-attention layer to obtain a context-aware feature $\mathcal{F}_c(\mathbf{x})\in\mathbb{R}^{d\times XY}$ from $\mathcal{F}_p(\mathbf{x})$. Then, the decoder part introduces a set of $N_o=100$ learnable object queries designed to represent a potential object within the image, allowing the model to focus on specific parts of the image relevant to individual objects. In this sense, the cross-attention stage utilizes object queries  $\mathcal{F}_c(\mathbf{x})$ to generate object representation vectors. In our configuration, we utilize 6 encoding and 6 decoding layers with 8 heads. Finally, each object representation can be used to predict $\mathcal{O}(\mathbf{x})$ and $\mathcal{L}(\mathbf{x})$, through feed-forward networks. For this process, we utilized a model in \cite{ref40}, which is trained on the COCO detection task. 

Obtaining $\mathcal{R}(\mathbf{x})$ from $\mathcal{F}(\mathbf{x})$ is fundamentally similar to the above method but requires a few distinct steps. First, it utilizes two separate branches for decoding, considering the fact that relation inherently involves two distinct objects and is sensitive to order; for example, \texttt{"man riding ski"} and \texttt{"ski riding man"} describe different relationships. In this sense, each branch utilizes distinctive $N_r=200$ learnable object queries $Q_{o1}\in\mathbb{R}^{d\times N_r}$ and $Q_{o2}\in\mathbb{R}^{d\times N_t}$. Also, different object embeddings $E_{o1}\in\mathbb{R}^{d\times N_r}$ and $E_{o2}\in\mathbb{R}^{d\times N_r}$ are added to each object query. Here, $N_r$ is the number of object pairs that are considered. Lastly, learnable relation embedding $E_r\in\mathbb{R}^{d\times N_t}$ is added to object queries on both branches. Thus, the resulting queries in each branch are expressed as: $Q_1 = Q_{o1} + E_{o1} + E_r$, $Q_2 = Q_{o2} + E_{o2} + E_r$. However, the object queries do not yet incorporate the context from the visual data. To address this, we concatenate these two queries and pass them through a self-attention layer. In this setup, we set $Q = K = [Q_1, Q_2]$, and $V = [Q_{o1}, Q_{o2}]$ where [, ] indicates the concatenation process. This encoding process allows the model to consider the relational context between the objects represented by each query. The output of the encoder is decoupled to continue the decoding process in each branch, and the following procedure is similar to the aforementioned process. Each branch obtains object representation through the cross attention with $\mathcal{F}_c(\mathbf{x})$ and utilizes feed-forward networks merging the outputs of two branches to predict $\mathcal{R}(\mathbf{x})$. Here, we utilize a model architecture with 6 encoding and 6 decoding layers, each with 8 attention heads. Finally, with a combination of $\mathcal{O}(\mathbf{x})$ and $\mathcal{R}(\mathbf{x})$, a scene graph $\mathcal{G}(\mathbf{x})$ is achieved. For this process, we adopted a model in \cite{ref33}, which is trained on the Visual Genome dataset. The object classes total 151, and the relation classes total 51. As depicted in Fig. \ref{fig:figure3}, by utilizing extracted semantic features i.e., $\mathcal{O},\mathcal{L},\mathcal{M}, \mathcal{R}$, and $\mathcal{G}$, a semantically aligned image can be generated by LDM following \cite{ref41}.

For feature map-based image generation, an iterative denoising process conditioned on $\mathcal{F}(\mathbf{x})$ might be inefficient in terms of both communicational burden and latency. Instead, a feature map $\mathcal{F}_d(\mathbf{x})\in\mathbb{R}^{C_d\times X_d\times Y_d}$ that can be directly decoded into the image can be used. So, we additionally adopt an encoder, i.e., a feature map extractor, which pairs with the decoder of the receiver \cite{ref42}.  Specifically the encoder consists of residual blocks which down-samples $\mathbf{x}$ into $\mathcal{F}_d(\mathbf{x})$ while decoder up-samples it and resulting $\mathbf{x}^*\in\mathbb{R}^{3\times M\times N}$. In our configuration, $X_d=Y_d=64$ and $C_d=4$. We freeze all the pretrained model parameters without requiring further training.

We emphasize that our key claim, i.e., task-adaptability, is not limited to the aforementioned model architecture. In ablations, we further confirm that our claim is shown to be valid for different architectures.

\textbf{Hyperparameters for Algorithm \ref{alg1} and \ref{alg2}:} Through exhaustive search, we found the thresholds in the algorithms for filtering scene graphs: \(\tau_f =\tau_r = 0.8\), respectively. These settings are found to preserve the essential semantic information while dramatically reducing data volume.
To filter less informative relations, we computed the probability of observing objects and relations using scene graph generation on 100,000 images from the Visual Genome dataset. 
 \subsection{Task Performance Evaluation}
Herein, two tasks are primarily considered: image generation and image retrieval, which are the most challenging ones.

For simpler tasks, classification is completed by sending $\mathcal{O}(\mathbf{x})$, localization is done by sending $\mathcal{L}(\mathbf{x})$, and detection is finished by transmitting a pair of $\mathcal{O}(\mathbf{x})$ and $\mathcal{L}(\mathbf{x})$ or semantic segmentation map $\mathcal{M}(\mathbf{x})$. Thus, we have excluded these simpler tasks from our analysis here, but we analyze them particularly in task completion time in Section \ref{subsec:latency}.

\textbf{Image Generation Task:} Fig. \ref{fig:figure3}, illustrates the process of image generation for Visual Genome.
We use two different metrics for evaluating image generation. As the performance metric for quantitative results, we use CLIP cosine similarity to measure semantic similarity at a fine-grained level \cite{ref43}. Fréchet Inception Distance (FID), which shows the overall similarity of the original and generated data distributions, is not suitable to our case, which assesses the sample-wise similarity. In contrast, a CLIP model utilizes a text-vision embedding space that captures high-level semantic information, making it more suitable for evaluating semantic alignment.

\begin{figure*}[t] 
    \centering
    \includegraphics[width=0.95\textwidth]{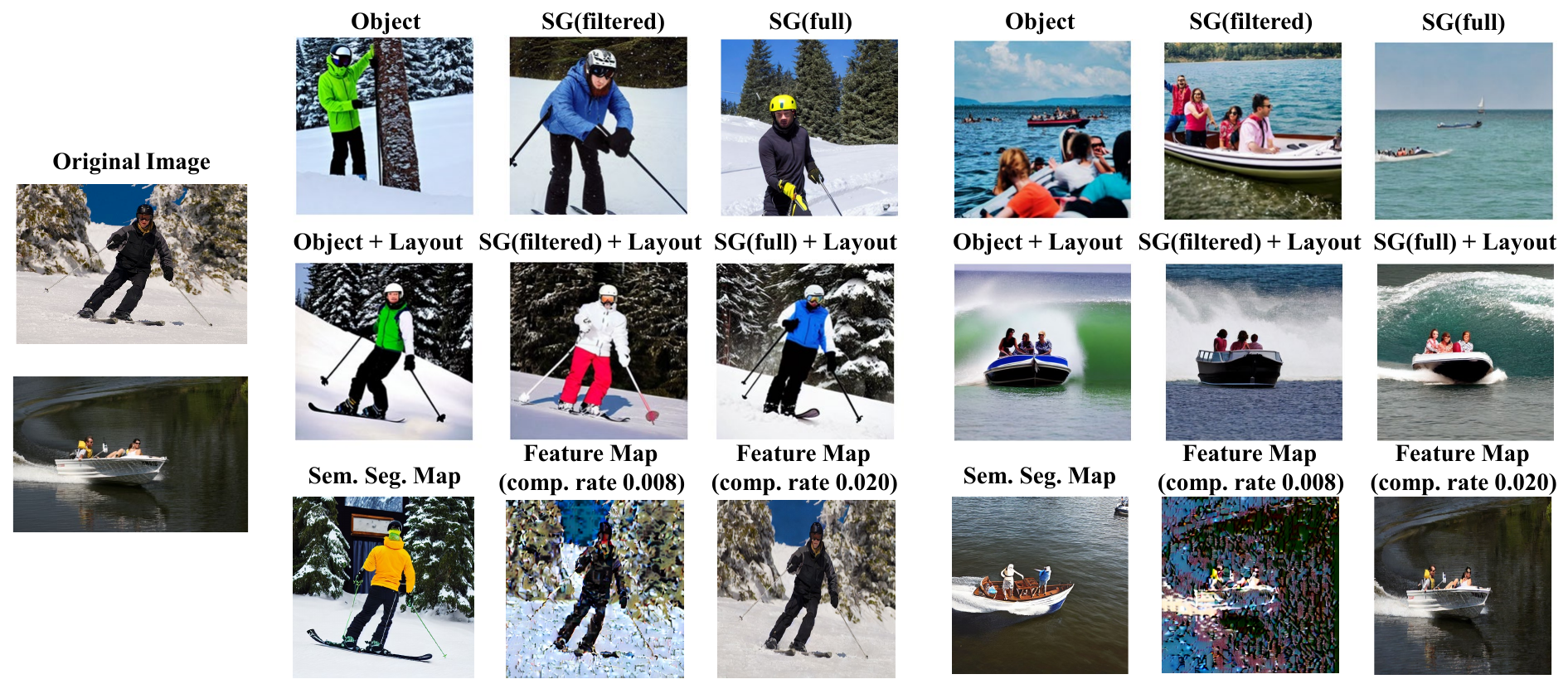}
    \caption{The qualitative results of the image generation task for Visual Genome. A given original image is regenerated by sending different visual semantics.}
    \label{gen_img_example}
\end{figure*}

For a pixel-wise perspective on how closely the generated image matches the original, we measure the structural similarity index (SSIM) \cite{ref44} and the peak signal-to-noise ratio (PSNR) between the two images. These are not a comparison at the semantic level, but reflect how the pixels of two images are similar.
To ensure consistent comparisons, both the original and generated images are resized to a fixed size of 224 $\times$ 224 before calculating these metrics. 
PSNR quantifies the ratio of the maximum possible pixel intensity to the mean squared error between the images, expressed in decibels.
SSIM evaluates perceptual similarity by comparing local patterns of luminance, contrast, and structural correlation between images, with values ranging from -1 to 1.

\begin{table}[t]
\centering
\caption{Performance of image generation on Visual Genome Dataset}
\begin{adjustbox}{width=0.48\textwidth}
\begin{tabular}{c|ccc}
\midrule[1.5pt]
\textbf{Semantic Features} & CLIP Similarity $\uparrow$ & SSIIM $\uparrow$ & PSNR $\uparrow$\\
\midrule[1.5pt]
Feature Map (comp. rate 0.020)  & $0.963$ & $0.867$ & $19.7$\\
%Feature Map (mid res.) \textcolor{red}{[45]} & $0.898$ & $50.1$ \\
Feature Map (comp. rate 0.008) & $0.806$ & $0.312$ & $13.4$\\
\midrule
SG + Layout & $0.856 $ & $0.159$ & $8.54$ \\
Filtered SG + Layout & $0.856$ & $0.159$ & $8.55$\\
Object + Layout & $0.855$ & $0.159$ & $8.51$\\
\midrule
Sem. Seg. Map & $0.841$ & $0.090$ & $8.84$\\
\midrule
SG & $0.832$ & $0.129$ & $8.39$\\
Filtered SG  & $0.832$ & $0.133$ & $8.38$\\
Object & $0.824$ & $0.133$ & $8.35$\\
\midrule[1.5pt]
\end{tabular}
\end{adjustbox}\label{tab1} \\
$\uparrow$: higher is better, SG: Scene Graph, i.e., $\mathcal{G}$, Filtered SG: Filtered Scene Graph, i.e., $\mathcal{G}_{\text{filtered}}$, comp. rate: Compression rate
\end{table}
Table \ref{tab1} presents the evaluation results of image generation tasks when using different visual semantics. Considering the randomness of the generating process due to the sampling process of the diffusion model, the results are averaged over testing 3,000 images. 
The results indicate that utilizing more diverse semantic information leads to higher performance metrics, with a particularly significant improvement observed when incorporating layout information. When comparing `Sem. Seg. Map' and `Object + Layout', despite `Sem. Seg. Map' delivers more detailed layout information that contributes to achieving lower pixel-wise error; it struggles to generate semantically similar images for scenes with complex arrangements. More importantly, our proposed semantic filtering algorithm shows minimal degradation, while the filtered scene graph (SG) uses approximately one-third of the full graph, demonstrating that the proposed algorithm effectively eliminates less informative and semantically redundant features. 
For SSIM and PSNR, it seems to say that the feature maps are the great choice for image generation, but it is not true when observing qualitative results.

\begin{figure}[t] 
    \centering
    \includegraphics[width=\columnwidth]{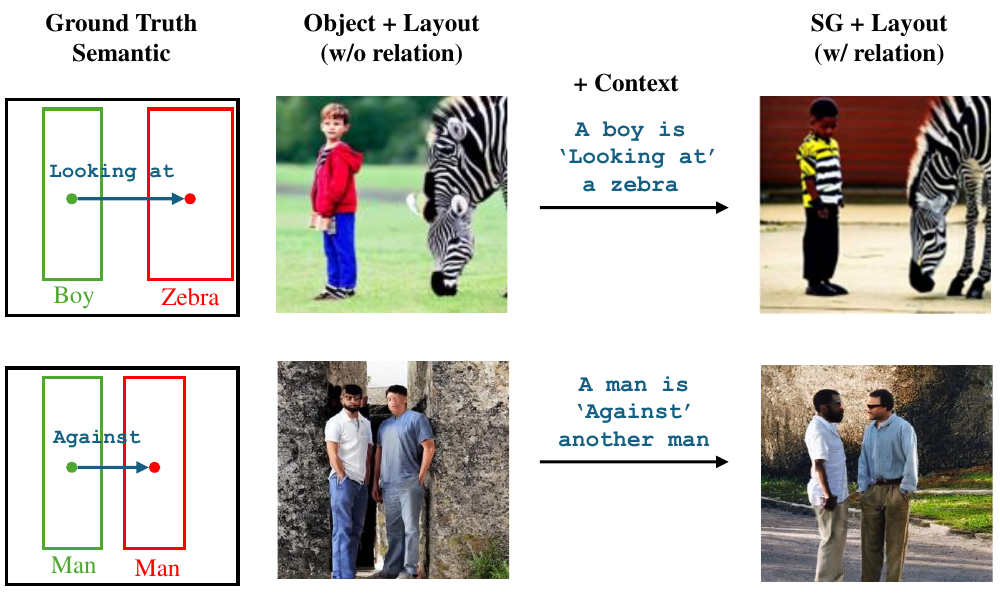}    \caption{The qualitative effects of the relations in the image generation task.}
    \label{relation_example}
\end{figure}

To provide a qualitative analysis, we include examples of the generated images in Fig. \ref{gen_img_example} to visualize the generated results from different semantics. 
Starting with the images generated using only object information, it is observed that while the objects appear in the scene, the generated images struggle to capture the context in cases where the relationships between objects are crucial. 
In contrast, as SG incorporates relation information, the generated images better reflect the contextual relationships among the objects, resulting in more coherent scenes. 
However, there is still a noticeable difference in the composition of the images. 
This issue is effectively addressed by adding simple layout information, such as bounding boxes. 
The three examples on the right in Fig. \ref{gen_img_example} show that incorporating layout information significantly enhances the semantic similarity of the generated images, allowing the model to produce images that are semantically much closer to the original. 
This improvement can be attributed to the fact that the combination of layout and object information allows the model to infer certain spatial relationships that correlate with the given relations, leveraging the statistical knowledge embedded in the generative model.

The two bottom-right examples illustrate the fidelity degradation of the `Feature Map' cases under compression. For the case of compression rate $0.008$, it exhibits the competitive pixel-level similarities (SSIM and PSNR), but it struggles to preserve high-level contextual coherence, leading to significantly lower CLIP similarity scores in Table \ref{tab1}. It coincides with the severely degraded fidelity in Fig. \ref{gen_img_example}.
With a light compression of $0.020$, it outperforms others, but requires extensive communication costs, which are to be discussed in Section \ref{subsec:latency}.
Conversely, the aforementioned approach—although not achieving superior pixel-level reconstruction—effectively conveys semantic content. These findings reveals that, depending on the underlying user's intention and the target performance metric, diverse types of semantics must be considered. Moreover, a much simpler semantics, such as filtered scene graphs, is sufficient for transmitting the semantically similar data.

Let us focus on the benefits of sending relations. 
Fig. \ref{relation_example} illustrates a case where the inclusion of relation information is crucial in generating semantically accurate images. 
The example in the figure shows where incorporating relation information helps capture the scene's context.

We emphasize that our semantic filtering block identifies the crucial relations that largely affect the visual semantics. Thus, the important relations are retained for transmission.

 \begin{figure}[t] 
    \centering
    \includegraphics[width=0.89\columnwidth]{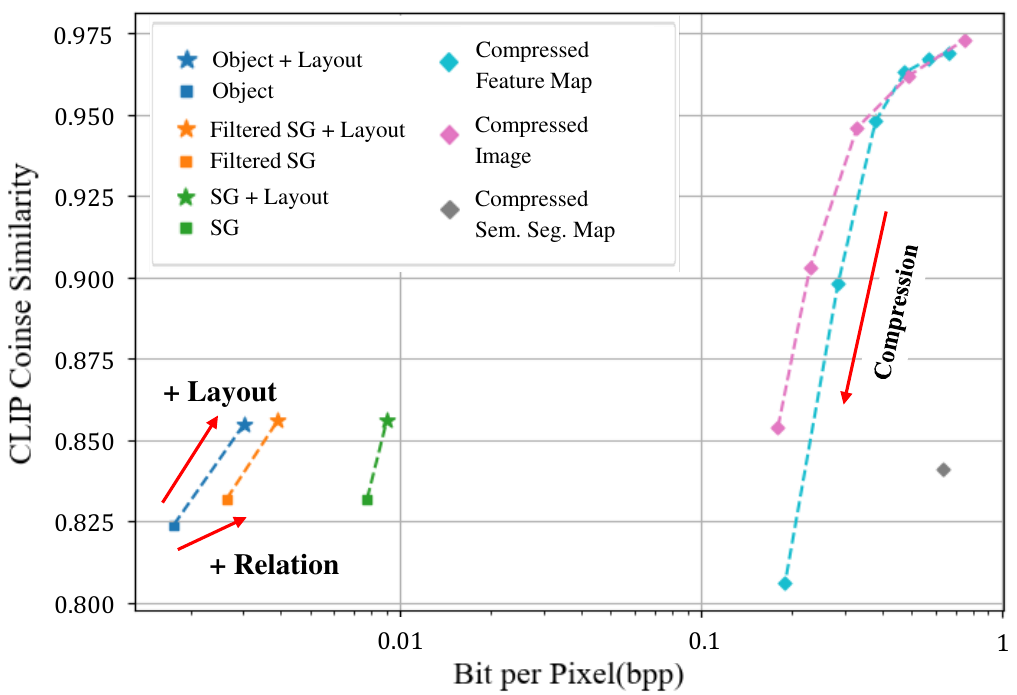}    \caption{CLIP cosine similarity v.s. bit per pixel for the image generation task}
    \label{Rate-Semantic Similarity}
\end{figure}

Fig. \ref{Rate-Semantic Similarity} illustrates the relationship between bit per pixel (bpp) and CLIP cosine similarity based on the type of semantic features. A smaller bpp means that fewer bits are required to transmit information. 
For `Compressed Image,' the conventional compression method, i.e., JPEG, is applied. For the feature map, we compressed it with different quantization levels, which is a commonly adopted method for feature map compression \cite{ref23}. The result shows that compressed images and feature map quickly degrade their similarity with a higher compression rate. 
When the bpp gets close to 0.1, `Compressed Image' demonstrates lower similarity than the 'Filtered SG + Layout' method when bpp goes under 0.18, while `Filtered SG + Layout' requires 45 times fewer bits than `Compressed Image.' On the other hand, `Compressed Feature Map' achieves lower similarity than `Object' at 0.19 bpp. The results emphasize that adaptive selection of visual semantics can significantly reduce the required transmission burdens.

Evaluation results for the Open Image V6 and ImageNet-1K datasets are presented in Section \ref{subsec:other_dataset}.

\begin{figure}[t] 
    \centering
    \includegraphics[width=\columnwidth]{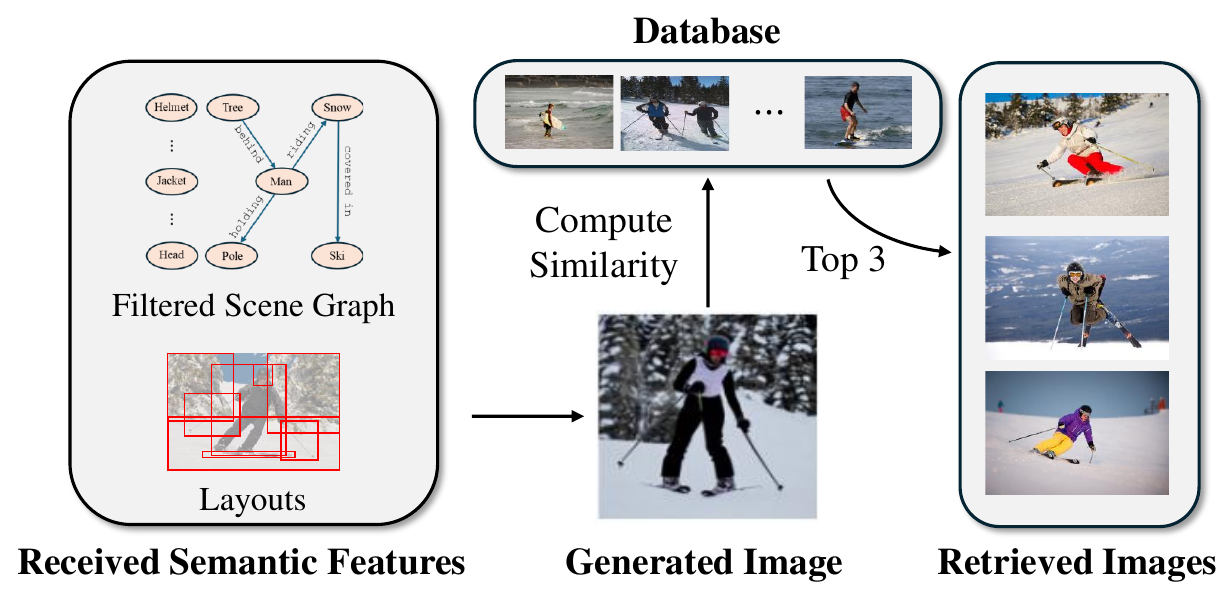}    \caption{Image retrieval task with a database for retrieving Top-3 images}
    \label{Image Retrieval Example}
\end{figure}

\textbf{Image Retrieval Task:} This task aims at finding images that are visually or semantically similar to a given query image from a pool of image samples called a `Database.'
In recent research, various methods have been employed for image retrieval, including feature hashing, deep metric learning, and graph-based retrieval algorithms. However, without loss of generality, we performed image retrieval using the CLIP cosine similarity between the generated image and the database samples. The overall flow of the task is depicted in Fig. \ref{Image Retrieval Example}. 
In our configuration, the database is composed of 10,000 images along with their corresponding embedding vectors.

We use a metric called `R@K (Recall at K)' for the performance measure, which evaluates how well the top K retrieved images match the relevant items in the dataset.
Precisely, R@K measures the proportion of relevant images that appear in the top K results out of the total number of relevant images. In our configuration, we have generated images using all or a subset of the semantic features extracted from the ground truth data. 
Then, we performed image retrieval using these generated images as queries. The retrieval is successful if the original ground truth image appears within the top K results. In this case, R@K performance is assigned a value of 1; otherwise, it is assigned a value of 0. 
It directly measures how effectively the retrieval aligns with the original semantics.

 \begin{table}[t]
\centering
\caption{Performance of image retrieval on Visual Genome Dataset}
\begin{adjustbox}{width=0.48\textwidth}
\begin{tabular}{c|cccc}
\midrule[1.5pt]
\textbf{Semantic Features}  & R@25 $\uparrow$ & R@10 $\uparrow$ & R@5 $\uparrow$ & R@1 $\uparrow$\\
\midrule[1.5pt]
Feature Map (comp. rate 0.020) & $0.944$ & $0.935$ & $0.922$ & $0.848$ \\
%Feature Map(mid res.) & $0.794$ & $0.713$& $0.646$ & $0.474$  \\
Feature Map (comp. rate 0.008)  & $0.111$ & $0.072$ & $0.054$ & $0.023$ \\
\midrule
SG + Layout  & $0.339$&  $0.237$ & $0.160$ & $0.053$ \\
Filtered SG + Layout  & $0.341$ &  $0.240$ & $0.167$ & $0.064$ \\
Object + Layout & $0.373$ & $0.232$ & $0.159$ & $0.049$ \\
\midrule
Sem. Seg. Map  & $0.221$ & $0.128$ & $0.075$ & $0.030$ \\
\midrule
SG (full)  & $0.192$ & $0.089$ & $0.059$ & $0.013$ \\
SG (filtered)  & $0.183$ & $0.084$ & $0.050$ & $0.012$\\
Object & $0.179$ & $0.089$ & $0.048$ & $0.006$\\
\midrule[1.5pt]
\end{tabular}
\end{adjustbox}\label{tab2} \\
$\uparrow$: higher is better, SG: Scene Graph, i.e., $\mathcal{G}$, Filtered SG: Filtered Scene Graph, i.e., $\mathcal{G}_{\text{filtered}}$, comp. rate: Compression rate
\end{table}

The performance results on 2,000 images are presented in Table \ref{tab2}.
The results show that using more diverse semantic features consistently leads to better retrieval performance, which coincides with the results from image generation tasks.
A significant improvement has been observed when layout information is included, indicating its substantial impact on enhancing retrieval performance. 
When comparing the performance between the full scene graph and the filtered scene graph with layouts, the filtered scene graph slightly but consistently outperforms the full version across all configurations.
This improvement can be attributed to providing the generative model with a more compact and refined set of semantic information, which aids the model in generating more relevant features for the retrieval process. Noteworthy, a simple set of semantics, i.e., `Object + Layout,' is best in a simple case with R@25, where we conjecture that the simple semantics seem proper in finding a wider range of similar images.
Also, the case of feature maps is severely degraded when the compression is adopted. We want to point out that the compressed feature map still suffers from large number of bits for transmission.

 \subsection{Demonstration on Wireless communication channels}\label{sec:4C}
To demonstrate the efficacy of our framework on wireless communication channels, we consider a Fifth-Generation New Radio (5G-NR) scenario, with the relevant system parameters summarized in Table \ref{tab3}. 
The configuration corresponds to NR numerology 0, and the low-density parity-check (LDPC) coding follows the specifications outlined in 3GPP TS 38.212.
For data with relatively large volume, e.g., images or feature maps, we employ longer code lengths to enhance reliability, whereas for data with smaller volume, we opt to use shorter code lengths to minimize latency.
We assume a situation where the user is allocated only two resource blocks (RBs), representing a scenario with limited communication resources.
We believe that efficient transmission via semantic communications becomes crucial in such resource-limited cases.

Fig. \ref{Throughputs} illustrates the number of visual semantics that can be transmitted per second over an additive white Gaussian noise (AWGN) channel with Rayleigh fading. The code rate is adapted to channel conditions, where lower rates are employed under worse channel conditions to ensure reliability.
For an in-depth comparison, we consider prior feature compression methods described in \cite{ref23, ref24}.
Each method can be described as follows: In \cite{ref24}, a semantic segmentation map $\mathcal{M}$ is decomposed into several one-hot encoded maps, and then JPEG is applied to them, i.e., only utilizing $\mathcal{L}$. We call it the `Compressed Semantic Segmentation Map.'
On the other hand, in \cite{ref23}, feature map $\mathcal{F}$ is masked on $\mathcal{M}$ then down-sampled, and the adjusted level of quantization is applied to it. We call it the `Masked Feature Map.' 
The simulation results demonstrate that representing semantics in a highly compact form can significantly reduce data transmission demands, showing the remarkable gap between the group of cases with images, semantic segmentation maps, and feature maps vs. the group of objects, layouts, and scene graphs. 
Specifically, `Filtered SG' shows 113 times larger throughput than `Masked Feature Map,' emphasizing that the compact visual semantics enable extremely faster transmission of visual information.
Moreover, our semantic filtering process can further accelerate the transmission, where `Filtered SG' shows 3 times larger throughput than `SG'.

For a naive transmission of images, merely one image transmission per second is achievable in low-SNR conditions for the compressed version, emphasizing that real-time transmission is infeasible.
In contrast, transmitting filtered scene graph and layout information, i.e., `Filtered SG + Layout,' enables sending hundreds of images per second even at 0 dB SNR. These results suggest that in scenarios with limited bandwidth, transmitting highly compressed semantic features is crucial for supporting real-time vision tasks.

\subsection{Estimation of End-to-End Latency}\label{subsec:latency}
\textbf{Latency Terms:}
Beyond confirming the reduced transmission time for our task-adaptive semantics in Sec. \ref{sec:4C}, 
the overall end-to-end latency, including the semantic encoding at the transmitter, transmission through the channel, and the completion of the visual task at the receiver, should be considered to verify the practicality of this system.
The resulting overall latency is obtained by summing all terms:
\begin{equation}
\tau_{\text{total}} \approx \tau_{\text{se}} + \tau_{\text{ce}} +  \tau_{\text{tx}} + \tau_{\text{cd}} + \tau_{\text{task}}.
\end{equation}

The first factor \(\tau_{\text{se}}\) represents the time required to extract visual semantics from an image, i.e., semantic encoding.
In our work, the semantic extractor relies on deep model architecture, e.g., ResNet-50 and auxiliary networks.
The latency encompasses the inference time of the semantic extractors.

The second factor \(\tau_{\text{ce}}\) is channel coding time, especially LDPC coding in our configuration. However, \(\tau_{\text{ce}}\) is negligibly small compared to other components, so we assume \(\tau_{\text{ce}} = 0\) during the latency measure.

\begin{figure}[t]
    \centering
    \includegraphics[width=\columnwidth]{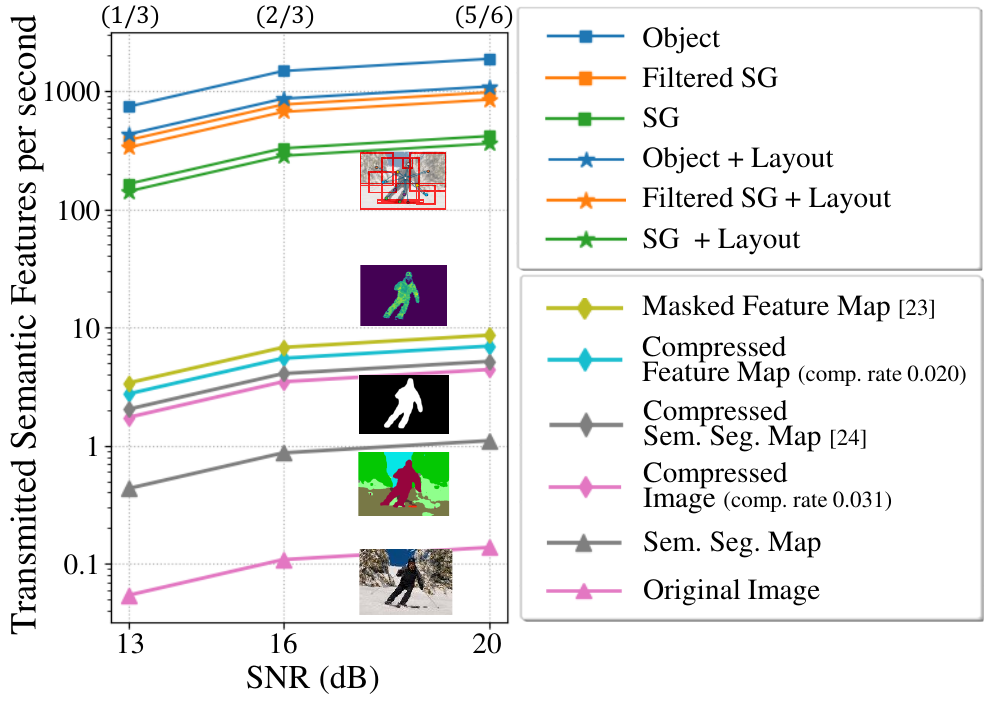}    \caption{The transmission throughput of different types of visual semantics on wireless communication channels, based on the Visual Genome Dataset}
    \label{Throughputs}
\end{figure}

The third factor \(\tau_{\text{tx}}\) is transmission latency, which is determined by channel conditions, available communication resources, and the compression level of semantic features. 
As shown in Fig. \ref{Throughputs}, different semantics show largely different transmission latency.

\begin{table}[t]
\centering
\caption{System Parameters of Wireless Communications Experiments}
\begin{adjustbox}{width=0.323
\textwidth}
\begin{tabular}{c|c}
\midrule[1.5pt]
\textbf{Parameters}  & \textbf{Values}\\
\midrule[1.5pt]
Channel bandwidth  & $5$ MHz\\
Subcarrier spacing  & $15$ KHz\\
Modulation & QPSK\\
\midrule
Source coding A$^*$ & JPEG (image)\\
Source coding B$^\dagger$ & ASCII (text)\\
\midrule
Channel model  & Rayleigh fading\\
Channel coding  & LDPC (base graph 1)\\
Codeword length A$^*$  & $8448$\\
Codeword length B$^\dagger$ & $1056$\\
Code rate & $(1/3)$, $(2/3)$, $(5/6)$\\
Decoding Iteration & 20\\
\midrule[1.5pt]
\end{tabular}
\end{adjustbox}\label{tab3}\\
$^*$:`Source coding A' and `Codeword length A' are for images, feature maps, and semantic segmentation maps. \\
$^\dagger$: `Source coding Type B' and `Codeword length  B' are for objects, relations, layouts, and scene graphs.
\end{table}

The fourth factor \(\tau_{\text{de}}\) is channel decoding time. Contrary to channel coding time, \(\tau_{\text{de}}\) can be substantially large as an LDPC code requires iterative decoding. However,  \(\tau_{\text{de}}\) can be significantly reduced when parallel computing is available. In other words, fast delivery of semantic features, i.e., small $\tau_{\text{se}} + \tau_{\text{ce}} +  \tau_{\text{tx}}$ can potentially reduce \(\tau_{\text{de}}\) per image.

The final factor \(\tau_{\text{task}}\) is the required time to complete the vision task based on the decoded data. This component can largely vary depending on the representation of the received data and the task type.
As described in Sec. \ref{sec:prelimB}, a group of simple tasks, such as classification, localization, and detection, can be completed right after receiving $\mathcal{O}$, $\mathcal{L}$, and $\mathcal{M}$ (or $\mathcal{O}$ plus $\mathcal{L}$), respectively, thus not requiring $\tau_{\text{task}}$. However, when the received data is, for example, a feature map, it requires semantic decoding and task operation based on an image, thus requiring $\tau_{\text{task}}$. 
For complicated tasks, such as image retrieval and generation, we need to run generative models, e.g., latent diffusion models, by utilizing visual semantics as the conditioning. All types of semantics, except for the conventional transmission of the image itself, require the same process.
After running generative models, the image generation task is terminated, but the retrieval task needs the search duration. Here, we exclude the search duration because it depends on the given database.

\textbf{Hardware Configurations:}
We utilized a single NVIDIA L4 GPU and a 12-core Intel Xeon CPU @ 2.20GHz, the common configuration of personal computing devices for running visual tasks. Herein, to account for the case where the transmitter lacks parallel computing resources, we also consider the cases where semantic extraction is done solely on the CPU. On the other hand, for complex tasks such as image generation, we utilized a single A100 GPU, which can provide sufficient computing power to complete the task.

\textbf{Detection Task Scenario:}
We first consider a scenario where a detection task is required on the receiver, as it can cover both classification and localization tasks. 
Here, we used the COCO dataset for the simulation.
For the comparison, we considered methods described in \cite{ref23, ref24}. Note that methods \cite{ref23, ref24} require semantic decoding, i.e., image generation for task completion, as their semantics are hard to use directly for completing detection tasks. 
Specifically, \cite{ref24} drops the object information from the semantic segmentation map; thus, reconstruction of objects is required, and \cite{ref23} utilizes the feature map; thus, the decoding procedure is indispensable. 

\begin{figure}[t]
    \centering
    \includegraphics[width=0.93\columnwidth]{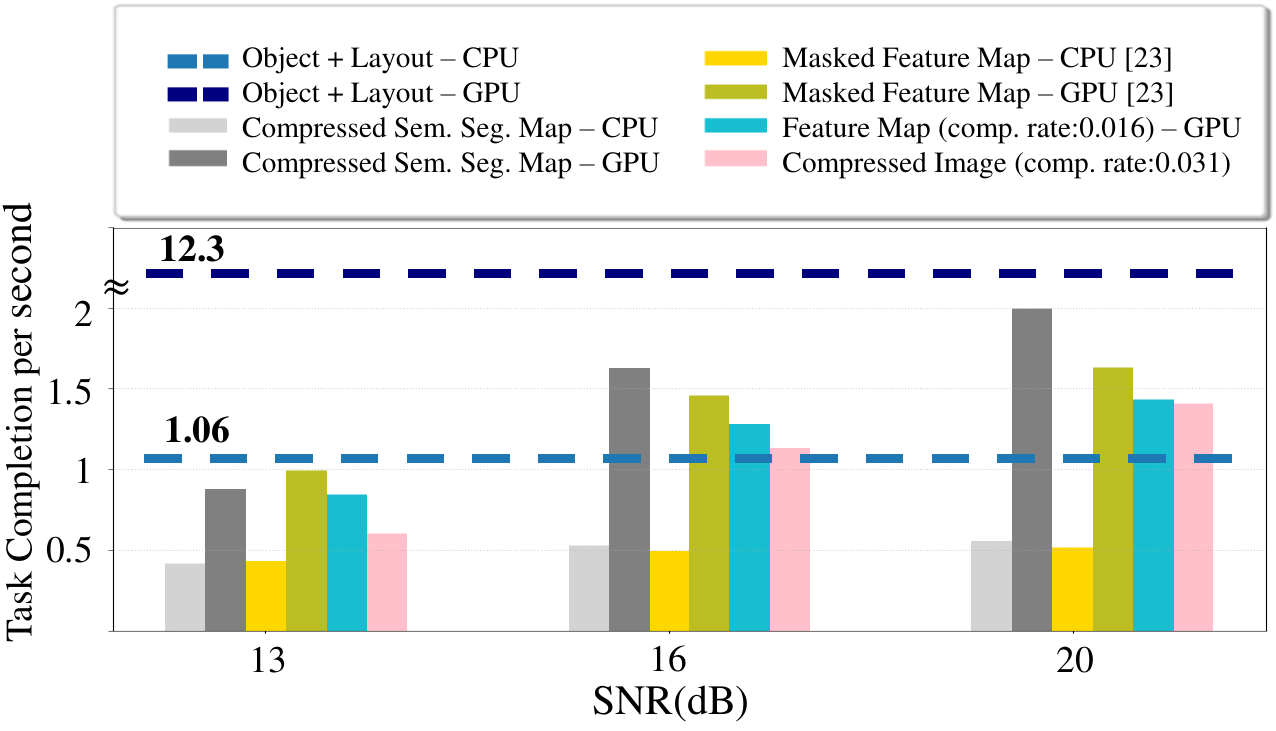}    \caption{Detection completion per second on different SNR regimes}
    \label{simple_task_latency}
\end{figure}

Fig. \ref{simple_task_latency} shows the number of tasks completed per second under varying channel conditions. For clarity, methods that showed higher latency than naively transmitting the original image are not included in the figure. For example, \cite{ref24} utilizes the pixel-level diffusion model \cite{ref21}, which consumes tens of seconds to generate a single image. 
For the semantic segmentation map-based method, we compressed it to the same level reported in \cite{ref24} by down-sampling width and height. For image-based task completion, we utilized a detection model that demonstrates low \(\tau_{\text{task}}\) \cite{ref40}. 

\begin{figure}[t] 
    \centering
    \includegraphics[width=0.97\columnwidth]{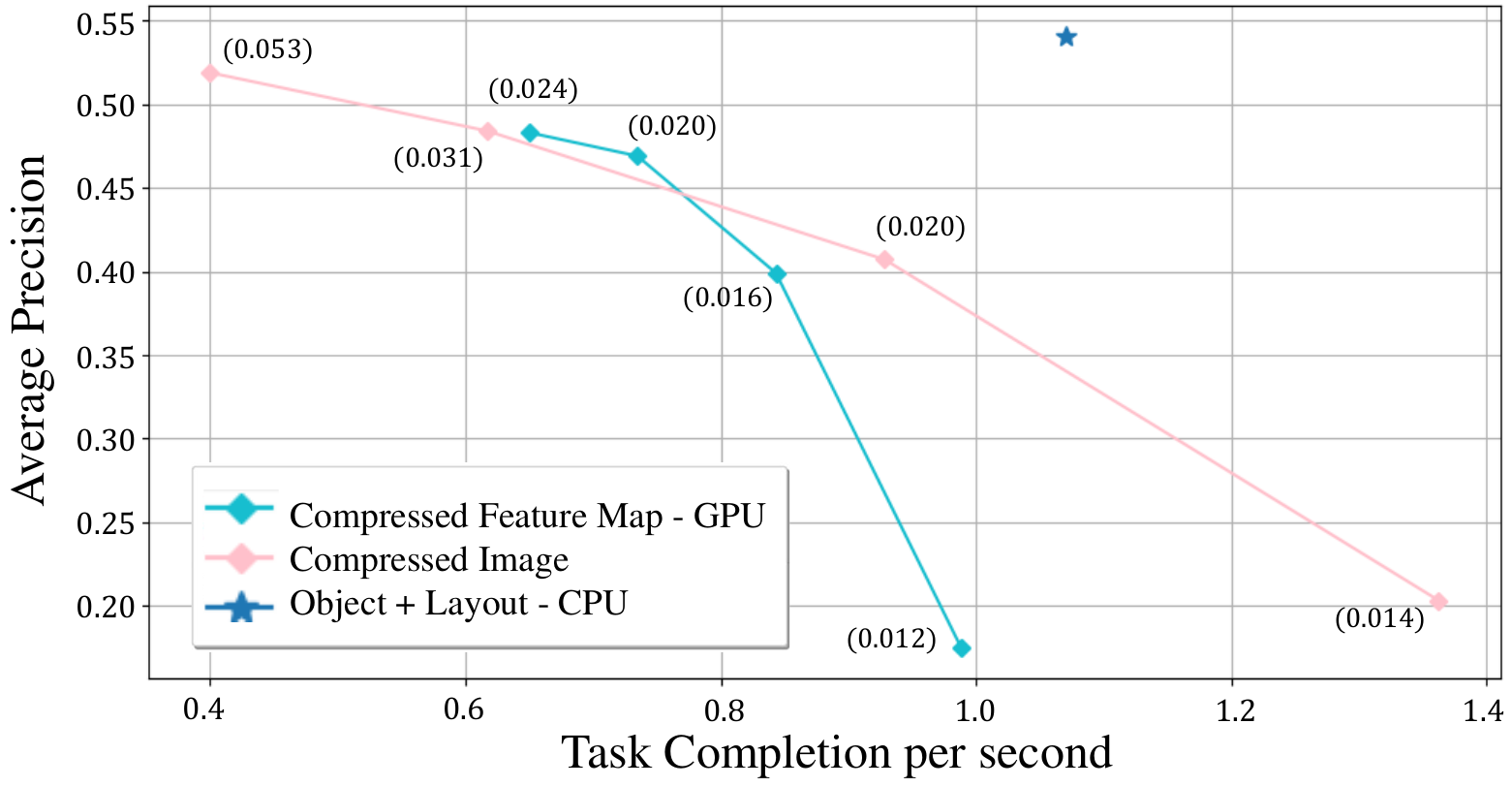}    \caption{Detection performance (average precision) v.s. task per second}
    \label{performance-latency}
\end{figure}

The simulation results in Fig. \ref{simple_task_latency} highlight the efficiency of our framework. With sufficient computing power at Tx, 12.3 tasks per second can be achieved, while other methods can complete under 2 tasks per second, even at a high SNR regime. On the other hand, when Tx lacks its computational power, the framework still can manage 1.06 tasks per second, outperforming all other methods based on CPU in all SNR regimes. 
More importantly, task operation on compressed images or generated images potentially degrades performance. To analyze the trade-off between compression level and task performance of each method, average precision (AP) degradation along higher compression levels is measured. Here, AP is a representative performance measure for detection tasks. Fig. \ref{performance-latency} shows the results tested on 1,000 COCO validation sets in 13dB SNR. The result shows that transmitting a compressed image can handle more images than `Object + Layout - CPU' only when largely sacrificing its performance. Comparing a compressed feature map and a compressed image, the former achieves higher performance at the same speed when less compressed, but it largely degrades its performance with higher compression. While we measured the latency of  \cite{ref24}, as it generates a lower resolution image than ours, we excluded it from measuring performance for fair comparison. Also, a segmentation map-based method can be easily expected to have a similar level of performance with `Object + Layout' as it entails the same semantic features, although it showed low speed as depicted in Fig.\ref{simple_task_latency}.

\textbf{Semantic Segmentation Task Scenario:}
Next, we evaluate a scenario in which Rx opts to execute semantic segmentation. We consider the Cityscapes test split, which comprises 500 images.
We compare our approach with two cutting-edge methods, including Elic \cite{ref45} and HiFiC \cite{ref46}, which are based on image reconstruction from received features via auto-decoder and generative adversarial networks, respectively.

\begin{figure}[t] 
    \centering
    \includegraphics[width=1\columnwidth]{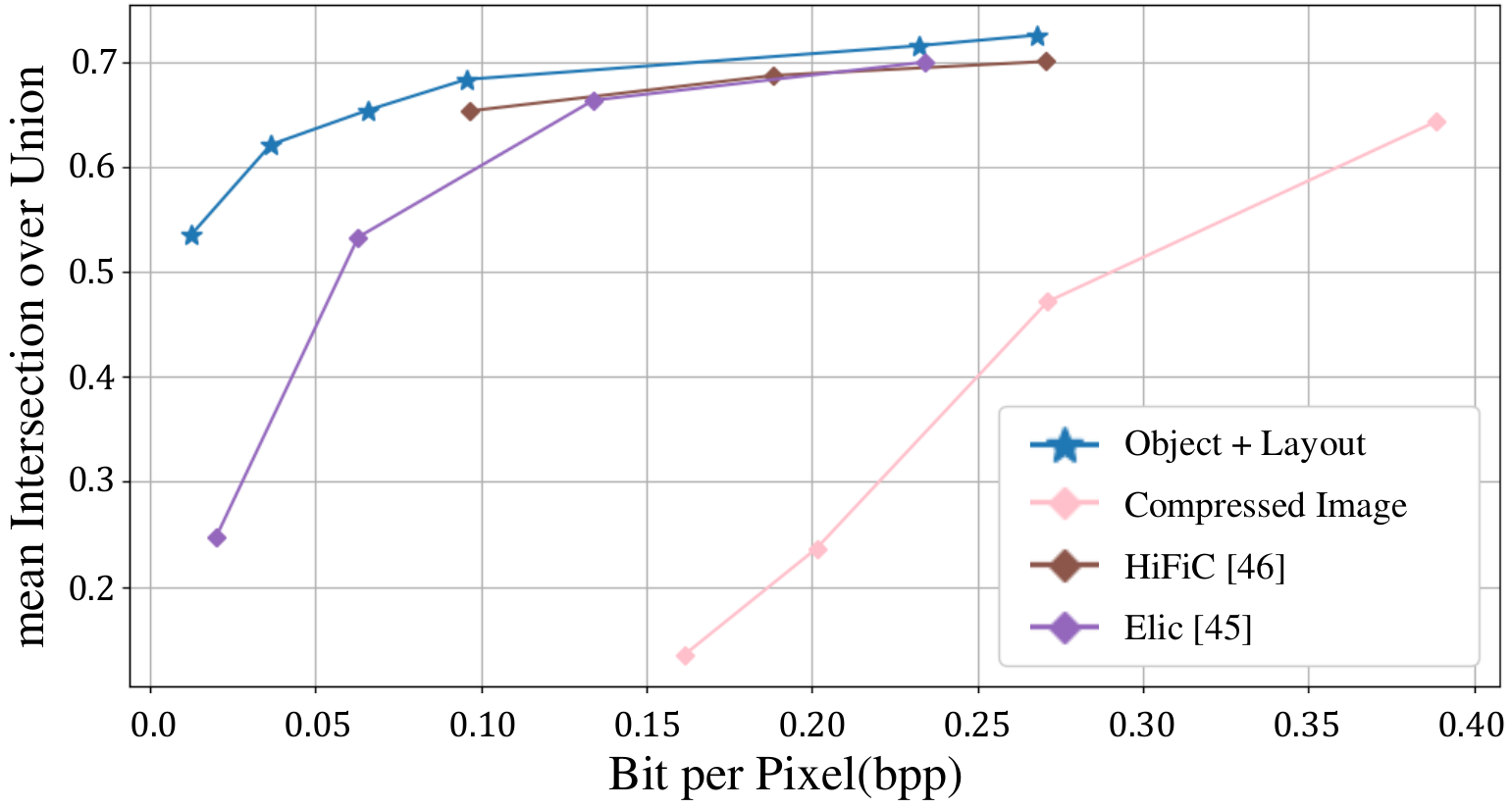}  \caption{Semantic segmentation performance v.s. bit per pixel}
    \label{seg-bpp-performance}
\end{figure}

\begin{figure}[t] 
    \centering
    \includegraphics[width=0.94\columnwidth]{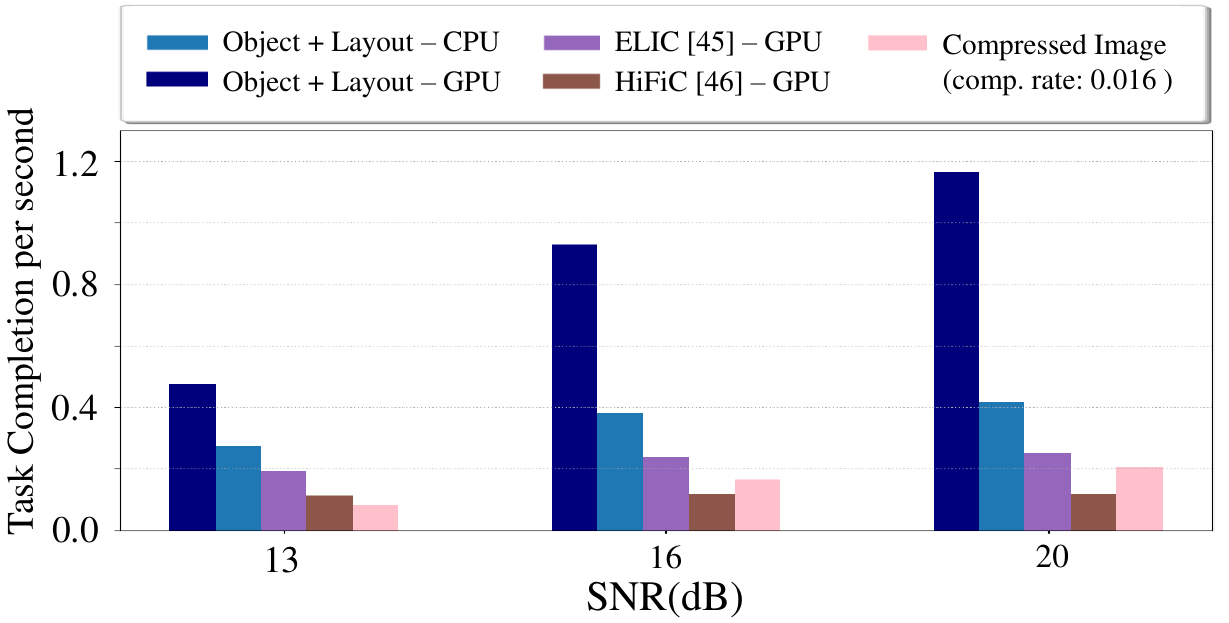}  \caption{Semantic segmentation completion per second on different SNRs}
    \label{seg-taskthroughput}
\end{figure}

Fig. \ref{seg-bpp-performance} presents mean Intersection over Union (mIoU) as a function of bpp. By following the task-adaptive semantic selection in Fig. \ref{fig:figure2}, we encode each object’s segmentation map as `Object + Layout' using run-length encoding and adjust the resolution of it to cover a range of bpp settings.
The results indicate that HiFiC, Elic, and image compression show degraded performance, coinciding with our observation in the detection task case. 
In contrast, `Object + Layout' outperforms others over the range of bpp, achieving mIoU of 0.536 at 0.013 bpp, whereas Elic shows only 0.247 mIoU at 0.020 bpp. 
Furthermore, HiFiC requires approximately twice the bpp of our approach to reach the mIoU of 0.68. These findings highlight the inefficiency of transmitting the semantic content of the current methods. 
Fig. \ref{seg-taskthroughput} shows the number of segmentation tasks completed per second under varying channel conditions, with all methods configured to achieve roughly 0.65 mIoU for fairness. 
These results underscore the critical importance of efficiently extracting task-relevant semantics with a task-optimized representation. Although Elic benefits from relatively fast inference, its throughput remains lower than that of the `Object + Layout — CPU', which utilizes limited computational resources. 
Also, our approach processes approximately 5.6 times more tasks than the `Compressed Image' baseline at 20 dB SNR.


\begin{figure}[t] 
    \centering
    \includegraphics[width=0.82\columnwidth]{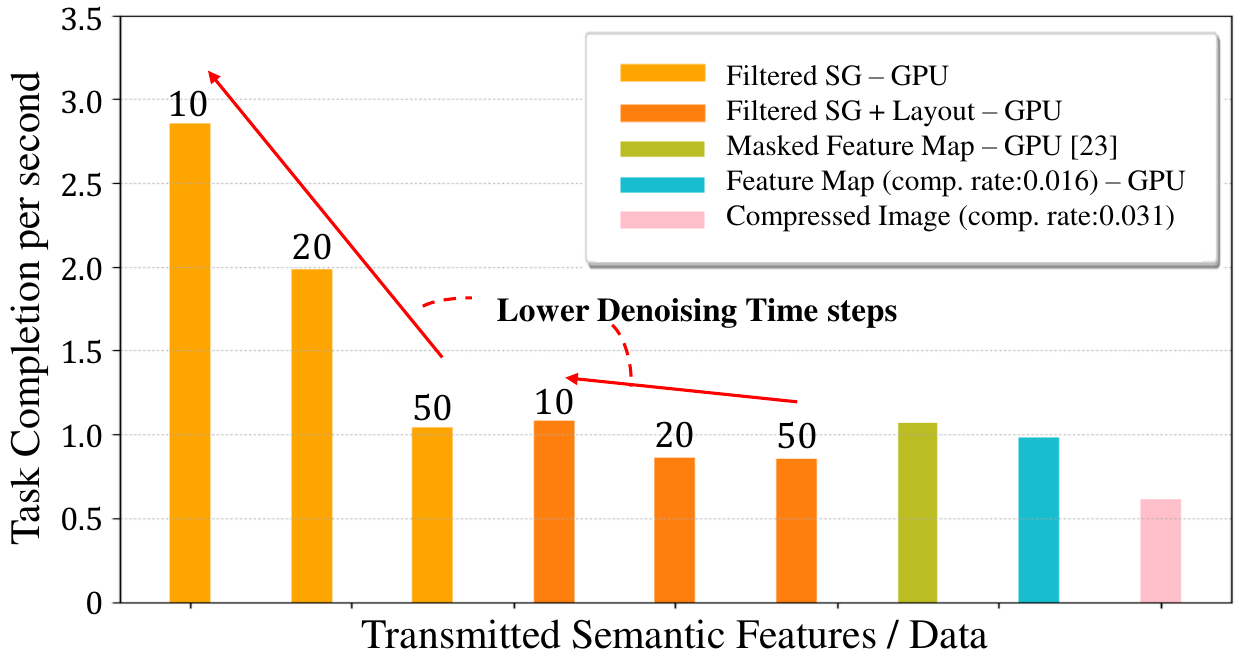}                \caption{Image generation completion per second for SNR $13$dB case}
    \label{generation_latency}
\end{figure}

\textbf{Complex Task Scenario:}
For more complex tasks such as image generation, simulation results are presented in Fig. \ref{generation_latency}. The results emphasize that the rapid delivery of semantic features is a critical factor impacting the number of task completions in a given time. Specifically, when generating a single image, using `Feature Map' and `Masked Feature Map' is significantly faster than employing `Filtered SG' or `Filtered SG + Layout,' which necessitates iterative denoising. 
However, in communication environments where various latency factors exist, it is observed that even with LDM, a higher number of tasks per second can be achieved. Especially lowering denoising time steps to 10, `SG - GPU' can deliver 4.6 times more images than `Compressed Image'.

\subsection{Evaluations on Other Datasets and Semantic Extractors}\label{subsec:other_dataset}
\textbf{Results on Open Image V6 and ImageNet-1K:} Table \ref{tab4} and Table \ref{tab5} present the image generation performance on Open Image V6 and ImageNet-1K, respectively. For each case, 3,000 images are used for image generation. 
As shown in \ref{tab4}, the result of Open Image V6 coincides with the findings in Visual Genome (referring to Table \ref{tab1}). Including layout information consistently yields gains, demonstrating its importance. Furthermore, the effectiveness of our semantic filtering is also evident; selectively transmitting only the most relevant semantics achieves minimal degradation in generation quality. 
For the ImageNet-1K results in Table \ref{tab5}, the inclusion of layout contributes relatively less than in other cases. 
We conjecture that this dataset focuses on a single-type object classification, where layout information is less critical. 

\begin{table}[t]
\centering
\caption{Performance of image generation on Open Image V6 Dataset}
\begin{adjustbox}{width=0.45\textwidth}
\begin{tabular}{c|ccc}
\midrule[1.5pt]
\textbf{Semantic Features} & CLIP Similarity $\uparrow$ & SSIIM $\uparrow$ & PSNR $\uparrow$\\
\midrule[1.5pt]
Feature Map (comp. rate 0.020)  & $0.986$ & $0.897$ & $28.8$\\
Feature Map (comp. rate 0.008) & $0.798$ & $0.332$ & $13.2$\\
\midrule
SG + Layout & $0.834 $ & $0.152$ & $8.60$ \\
Filtered SG + Layout & $0.834$ & $0.152$ & $8.60$\\
Object + Layout & $0.833$ & $0.152$ & $8.53$\\
\midrule
Sem. Seg. Map & $0.787$ & $0.097$ & $9.42$\\
\midrule
SG & $0.810$ & $0.116$ & $8.30$\\
Filtered SG  & $0.809$ & $0.123$ & $8.38$\\
Object & $0.806$ & $0.122$ & $8.39$\\
\midrule[1.5pt]
\end{tabular}
\end{adjustbox}\label{tab4} \\
\end{table}
\begin{table}[t]
\centering
\caption{Performance of image generation on ImageNet-1K Dataset}
\begin{adjustbox}{width=0.45\textwidth}
\begin{tabular}{c|ccc}
\midrule[1.5pt]
\textbf{Semantic Features} & CLIP Similarity $\uparrow$ & SSIIM $\uparrow$ & PSNR $\uparrow$\\
\midrule[1.5pt]
Feature Map (comp. rate 0.020)  & $0.979$ & $0.865$ & $28.5$\\
Feature Map (comp. rate 0.008) & $0.798$ & $0.316$ & $12.7$\\
\midrule
SG + Layout & $0.819 $ & $0.138$ & $8.56$ \\
Filtered SG + Layout & $0.818$ & $0.139$ & $8.50$\\
Object + Layout & $0.814$ & $0.139$ & $8.44$\\
\midrule
Sem. Seg. Map & $0.778$ & $0.102$ & $8.84$\\
\midrule
SG & $0.804$ & $0.130$ & $8.80$\\
Filtered SG  & $0.803$ & $0.134$ & $8.80$\\
Object & $0.798$ & $0.134$ & $8.76$\\
\midrule[1.5pt]
\end{tabular}
\end{adjustbox}\label{tab5} \\
\end{table}

\textbf{Results with Other Semantic Extractors:} To validate with other semantic extractors, we additionally test HRNetW32 + CNN in \cite{ref47} and VGG16 + LSTM in \cite{ref48} for the image generation task on Visual Genome. We here run `SG' and `Filtered SG' with varying extractors. Our methods demonstrate sufficiently high CLIP similarities across different extractors, validating that our approach is applicable to various model architectures.
 \begin{table}[t]
\centering
\caption{Performance of image generation with Different Semantic Extractors on Visual Genome}
\begin{adjustbox}{width=0.43\textwidth}
\begin{tabular}{c|cccc}
\midrule[1.5pt]
\textbf{Semantic Extractor} & CLIP Similarity $\uparrow$ & SSIIM $\uparrow$ & PSNR $\uparrow$ &Latency $\downarrow$\\
\midrule[1.5pt]
ResNet50 + Transformer & $0.832/0.856$ & $0.129/0.159$ & $8.39/8.54$ & $30ms$\\
HRNetW32 + CNN\cite{ref47}  & $0.816/0.835$ & $0.132/0.136$ & $8.63/8.64$ & $60ms$\\
VGG16 + LSTM\cite{ref48} & $0.817/0.842$ & $0.127/0.157$ & $8.54/8.82$ & $80ms$\\
\midrule[1.5pt]
\end{tabular}
\end{adjustbox}\label{tab6} \\ -/- means the performance of `Filtered SG'/`Filtered SG + Layout'. \\ $\uparrow$: higher is better. $\downarrow$: lower is better. `Latency' indicates $\tau_{se}$
\end{table}

\subsection{Summary of Key Principles}
Our experimental results substantiate the critical role of task-adaptability and semantic filtering in real-time semantic communication systems. The principal findings may be distilled into the following key points:
\begin{itemize}
\item \textbf{Task-Adaptive Feature Selection:} Our approach is particularly crucial in a bandwidth-limited environment. For the image generation task on the Visual Genome dataset, transmitting `Filtered SG + Layout' has achieved higher semantic similarity than a compressed image, requiring 45 times fewer bits. On the other hand, for the semantic segmentation task on the Cityscape dataset, transmitting `Object + Layout' has shown higher mIoU than a compressed image, requiring 5.9 times fewer bits.
\item \textbf{Effective Filtering of Semantic:} Effective semantic compression can remain robust against task performance degradation. Our filtering block reduces the bits required to transmit the scene graph to one-third of the original amount while incurring minimal performance degradation for the image generation task. Under the Cityscapes dataset semantic segmentation task, `Object + Layout' has shown 0.536 mIoU even under 0.013 bpp, while a competitive baseline, i.e., Elic, requires 5 times more bits to achieve similar performance.
\item \textbf{Reduced End-to-End Latency:} Our approach can be directly applied to performing tasks, which further reduces end-to-end latency. 
For the detection task on the COCO dataset, `Object + Layout' could complete 8.8 times more tasks than `Feature Map' within a given time.
\end{itemize}

 \section{Conclusion}\label{sec:conclusion}
In this paper, we suggest an efficient and effective semantic communication system for real-time CV tasks with the consideration of three key design points: \textbf{i)} Capability to select semantic features to transmit, \textbf{ii)} Filtering less informative semantic features to reduce transmitted bits, \textbf{iii)} Efficiently utilizing the received semantic features to perform the task. 
In this sense, we thoroughly examine how much efficiency can be achieved by using compact visual semantics and how well the given CV tasks can be completed.
Based on this study, we have analyzed the task performance for both simple and complicated tasks, demonstrating the impact of selective feature transmission on task performance. 
Our simulation results reveal that even with a minimal bit representation, compactly expressed semantic features can achieve sufficient task completion. Also, the simulations on limited bandwidth validate that task-adaptive selection of visual semantics is a possible enabler for running real-time CV tasks.

\bibliographystyle{IEEEtran}
\bibliography{references}



\begin{IEEEbiography}[{\includegraphics[width=1in,height=1.25in,clip,keepaspectratio]{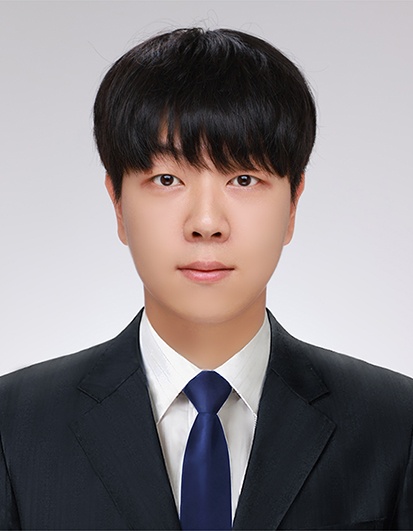}}]{Jeonghun Park}
received his B.S. and M.S. degrees in Electronics Engineering from Pusan National University, Pusan, South Korea, in 2021 and 2024, respectively. Since 2024, he has been pursuing a Ph.D. degree at the Graduate School of Artificial Intelligence, Ulsan National Institute of Science and Technology (UNIST), Ulsan, South Korea. His research interests include 6G intelligent communication and semantic communication.
\end{IEEEbiography}
\begin{IEEEbiography}
[{\includegraphics[width=1in,height=1.25in,clip,keepaspectratio]{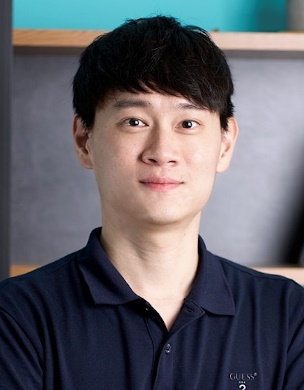}}]{Sung Whan Yoon}
received his B.S., M.S., and Ph.D. degrees in the Department of Electrical Engineering
from Korea Advanced Institute of Science and Technology (KAIST), Daejeon, South Korea, in 2011, 2013, and 2017, respectively. He was a postdoctoral researcher at KAIST from 2017 to 2020. He is currently working as an Associate Professor at the Artificial Intelligence Graduate School (AIGS) and the Department of Electrical Engineering at Ulsan National Institute of Science and Technology (UNIST), Ulsan, South Korea. 
His research interests are in machine learning with a special focus on i) learning theory to reveal the principles of deep training via a lens of information theory and signal processing, and ii) deep algorithmic methods to bolster the generalization of learners across diverse settings. 
For another direction, he has dedicated himself to developing intelligent communication systems beyond Shannon's theory. 
\end{IEEEbiography}
\vspace{11pt}
\vfill
\end{document}